\documentclass[showpacs,preprintnumbers]{revtex4}%
\usepackage{amsfonts}
\usepackage{amsmath}
\usepackage{graphicx}
\usepackage{dcolumn}
\usepackage{bm}
\usepackage{amssymb}
\usepackage[latin1]{inputenc}
\usepackage[gen]{eurosym}
\usepackage{dcolumn}
\usepackage{color}
\usepackage{fancybox}
\usepackage{amsbsy}
\usepackage{subfigure}%
\DeclareMathAlphabet{\mathpzc}{OT1}{pzc}{m}{it}
\begin{document}
\title{The $O(2)$ model in polar coordinates at nonzero temperature}
\author{M. Grahl$^{\text{(a)}}$, E. Seel$^{\text{(a)}}$, F.
Giacosa$^{\text{(a)}}$, and D. H.\ Rischke$^{\text{(a,b)}}$}
\affiliation{$^{\text{(a)}}$Institute for Theoretical Physics, Goethe University,
Max-von-Laue-Str.\ 1, D--60438 Frankfurt am Main, Germany }
\affiliation{$^{\text{(b)}}$Frankfurt Institute for Advanced Studies, Goethe University,
Ruth-Moufang-Str.\ 1, D--60438 Frankfurt am Main, Germany }

\begin{abstract}
We study the restoration of spontaneously broken symmetry
at nonzero temperature in the framework of the $O(2)$ model 
using polar coordinates. We apply the CJT
formalism to calculate the masses and the condensate
in the double-bubble approximation, both with and without
a term that explicitly breaks the $O(2)$ symmetry.
We find that, in the case with
explicitly broken symmetry, the mass of the angular degree
of freedom becomes tachyonic
above a temperature of about $300$ MeV.
Taking the term that explicitly breaks the symmetry to be
infinitesimally small, we find that the Goldstone theorem 
is respected below the critical temperature. However, this
limit cannot be performed for temperatures above the 
phase transition.
We find that, no matter whether we break the symmetry explicitly
or not, there is no region of temperature in which the radial
and the angular degree of freedom 
become degenerate in mass.
These results hold also when the mass of the
radial mode is sent to infinity.

\end{abstract}

\pacs{11.10.Wx,12.39.Fe,11.30.Rd}
\maketitle





\collaboration{MUSO Collaboration}

\collaboration{CLEO Collaboration}

\section{Introduction}

The $O(N)$ model is of great importance for the theory of critical phenomena
\cite{Kleinert:2001ax} and has been intensely studied in different theoretical
frameworks
\cite{MeyersOrtmanns:1993dw,Roh:1996ek,Baacke:2002pi,Andersen:2004ae,Petropoulos:2004bt,Ivanov:2005yj,Li:2009by,Argyres:2009sw}.
Using a mass term with a `wrong' sign, the $O(N)$ symmetry is
spontaneously
broken to $O(N-1)$ in the vacuum. At sufficiently high temperature $T$, the
$O(N)$ symmetry is restored. The
case $N=4$ has been profusely investigated because of its application to
the chiral phase transition in quantum chromodynamics (QCD)
\cite{Pisarski:1983ms,Wilczek:1992sf,Lenaghan:1999si,Butti:2003nu,MeyerOrtmanns:2007qp,Seel:2011ju}.
QCD has a chiral $SU(N_f)_R \times SU(N_f)_L$
symmetry (where $N_f$ is the number of quark flavors)
which is spontaneously broken to $SU(N_f)_{R+L}$ in
the vacuum. The group $O(4)$ is locally isomorphic to $SU(2)\times
SU(2)$, and $O(3)$ is locally isomorphic to $SU(2)$.
Therefore, using the universality hypothesis, the $O(4)$ model can 
be considered as an effective theory for the restoration of
chiral symmetry in QCD with two quark flavors. 

Most studies of symmetry restoration within the $O(N)$ model
\cite{Pisarski:1983ms,Wilczek:1992sf,Lenaghan:1999si,MeyerOrtmanns:2007qp,Seel:2011ju}
have been performed in Cartesian coordinates, $\vec{\phi} = (\phi_1,\,
\phi_2, \, \ldots, \phi_N)$. Then, when the symmetry is spontaneously
broken, the first coordinate is usually
selected to assume a non-vanishing vacuum expectation value $\varphi$,
$\phi_1 \rightarrow \varphi + \phi_1$, and the remaining $N-1$
coordinates are taken to be the $N-1$ Goldstone modes arising from
spontaneous symmetry breaking. However, in the absence of explicit
symmetry-breaking terms, the effective potential is of the
well-known `Mexican hat'-type, with any state on
the circle $|\vec{\phi}|= \varphi = const.$ being energetically
degenerate with the vacuum state $\varphi$. Therefore, the
Goldstone modes actually correspond to angular variables 
when moving along the circle
$|\vec{\phi}| = const.$, and the massive degree of freedom 
corresponds to the radial degree of freedom when trying to `climb' the
rim of the hat. 

This observation constitutes the main motivation for
the present paper where we study symmetry restoration within
the $O(N)$ model in polar coordinates.
For the sake of simplicity, we shall concentrate on the case 
$N=2$, because the transition from Cartesian to polar coordinates
is particularly simple in this case. Nevertheless, the
conceptual issues are the same in this case as for $N=4$. In a certain
sense, the case $N=2$ corresponds to QCD with one flavor in the 
absence of the axial anomaly. For $N_f=1$, the (most simple) anomaly term would
introduce a linear term in the massive degree of freedom,
corresponding to an explicit symmetry breaking term. This is
different from the two-flavor case, i.e., $N=4$, which corresponds to
the case of maximal anomaly or maximal $U(1)_A$ breaking.
Due to the similarities with the chiral symmetry of QCD and its breaking
and restoration at nonzero $T$, in the following 
we shall also refer to the transition
in the $O(2)$ model as `chiral transition'. The order parameter
for the transition, $\varphi$, will be called `chiral condensate', 
and the two degrees of freedom (the `chiral partners') are referred to
as the $\sigma$ meson and the pion. 
The so-called `chiral limit' is the case where explicit symmetry-breaking
terms are sent to zero, in which case the pion becomes a
true Goldstone boson.

Another motivation for our work is that, if we send the mass of
the radial degree of freedom (the $\sigma$ meson) to infinity, 
we just obtain the leading-order Lagrangian of 
chiral perturbation theory ($\chi PT$)
\cite{Gasser:1983yg,Ecker:1988te,Scherer:2002tk}, with the
Goldstone boson (the pion) as angular degree of freedom. 
In our simplified framework of the $O(2)$ model,
it is then possible to perform a test of the validity of 
angular variables at nonzero $T$.

The transition from Cartesian to polar coordinates represents a change of the
representation of the same theory. Obviously, physical quantities should be
independent from the adopted representation. This fact is ensured by
the so-called $S$-matrix equivalence theorem \cite{Kamefuchi:1961sb} 
which states that the elements of the $S$-matrix do not change when 
performing a transformation of the fields. However, for actual
calculations one must use a certain approximation scheme.
Then, the results obtained in one representation are not necessarily equal to
those obtained in another representation \cite{Kondratyuk:2001bw}. 
In this work we shall explicitly show that, by using the CJT formalism
in the double-bubble approximation
\cite{Cornwall:1974vz}, quantities computed in polar
coordinates are actually quite different from those evaluated in
the standard Cartesian coordinates.

Our results are the following: (i) for explicit
symmetry breaking, the pion mass becomes tachyonic at
high $T$, signalizing a breakdown of the
CJT formalism in the double-bubble approximation. (ii) When
the explicit symmetry-breaking term is sent to zero (which, in analogy
to the case $N=4$, we refer to as the chiral limit), the Goldstone
boson (pion) remains massless at each $T$ in polar coordinates, thus
satisfying the Goldstone theorem. This is contrary to Cartesian coordinates
where a nonzero value of the pion mass is obtained as soon as the temperature
is switched on, see e.g.\ Ref.\ \cite{Lenaghan:1999si}. However, 
due to singular terms $\sim 1/\varphi$ in the equations for the masses
and the condensate, and
because the order parameter $\varphi = 0$ above the phase transition,
the model becomes ill-defined.
(iii) Both with and without explicit symmetry breaking, there
is no region of temperature in which the chiral partners become
degenerate in mass. (iv) Similar results hold also in the nonlinear
limit, i.e., when the radial excitation becomes infinitely heavy.

The reason why polar coordinates are problematic at high $T$ can be traced
back to the decreasing value of the chiral condensate $\varphi$: in fact, in
polar coordinates there are interaction terms proportional to inverse powers
of $\varphi$, which render the application of the CJT formalism (or any other
resummation scheme) problematic when $\varphi$ is too small. In order to
circumvent this problem one can perform a slightly different transformation to
polar coordinates, in which the polar coordinates are not defined with respect
to the origin of the Cartesian coordinates \cite{Argyres:2009em}. In this way a smooth limit from
Cartesian to polar coordinates is realized, in which all the results of
Cartesian coordinates can be reobtained. Interestingly, it is possible to
investigate these issues also in the very simple situation of a free
Lagrangian, see Sec.\ \ref{free} for details. In the Cartesian representation
the results for thermodynamical quantities, such as the pressure, are exact in
this case. The deviations of the results in the polar representation from the
Cartesian one explicitly show the limitations of polar coordinates.

A further issue of polar coordinates is the fact that the Jacobian associated
with the field transformation is not unity: an additional term emerges in the
transformation of the interaction measure, which is potentially relevant in
the context of quantum field theory. We discuss in detail why it is justified to 
neglect its contribution for our conclusions. For this purpose, we
rely on perturbative 
cancellations between Jacobian contributions and certain divergences
(which will be shown in detail in Appendix \ref{cancel}), we discuss 
the vanishing of the Jacobian contributions in the dimensional regularization scheme, and
we study a different field representation in terms of polar variables, in which
the Jacobian is indeed unity. In all cases we studied, the qualitative
picture does not change and the
same conceptual issue of diverging interaction terms in the high-temperature
region exists also in this case.

The paper is organized as follows: In Sec.\ \ref{sec} we write the
$O(2)$ model in terms of polar coordinates and discuss the subtleties concerning
this coordinate transformation. In Sec.\ \ref{res} we apply the CJT
formalism and present the numerical results. The simple case of a free
Lagrangian is discussed in Sec.\ \ref{free}. In Sec.\ \ref{elina2} the
alternative representation with unit Jacobian is discussed. Finally, 
we give our conclusions and an outlook in Sec.\
\ref{conclusionsoutlook}.
Our units are $\hbar = c = k_B = 1$; the metric tensor is $g_{\mu \nu}
= {\rm diag} \, (+,-,-,-)$.

\section{The $O(2)$ model in polar coordinates}

\label{sec}

\subsection{Tree level, zero temperature}

\label{sec1}

The $O(2)$ model in Cartesian coordinates
$\vec{\phi}=(\phi_{1},\phi_{2})$, including an explicit
symmetry breaking term $\sim H$, is described by the Lagrangian
\begin{equation}
\mathcal{L}_{cart}=\frac{1}{2}\partial_{\mu}\vec{\phi} \cdot
\partial^{\mu}\vec{\phi}+\frac{m^{2}}{2}\vec{\phi}\cdot
\vec{\phi}-\frac{\lambda}{2}(\vec{\phi}
\cdot \vec{\phi})^{2}+H\phi_{1}\; .\label{lcart}%
\end{equation}
As usual, we consider the shift $\phi_{1}\rightarrow\phi_{1}+\phi,$ where
$\phi$ is a constant. At zero temperature the minimization of the potential
$V(\phi_{1}=\phi,\phi_{2}=0)$ leads to the minimum $\phi=\varphi$
satisfying the following equation:
\begin{equation}
m^{2}\varphi-2\lambda\varphi^{3}+H=0\text{ .}\label{fix1}%
\end{equation}
When $H=0$ the symmetry of the Lagrangian (\ref{lcart})
under $O(2)$ transformations is exact. For $m^{2}>0$
and $H=0$ the global minimum is realized for $\phi\equiv\varphi\neq 0$, i.e.,
the ground state breaks the $O(2)$ symmetry spontaneously. The
vacuum expectation value $\varphi$ is referred to as `chiral condensate'
in our model. In the vacuum, the
numerical value for $\varphi$ is chosen to be the pion decay
constant, $\varphi\equiv f_{\pi}=92.4$ MeV. When $H\neq0$ an additional
explicit breaking of chiral symmetry is realized.

By shifting the fields around their values at the global minimum, one obtains
the zero-temperature tree-level masses $m_{1}$ and $m_{2}$ as
\begin{equation}
m_{1}^{2}=-m^{2}+6\lambda f_{\pi}^{2}\ ,\ \ m_{2}^{2}=-m^2+2\lambda
f_{\pi}^2=\frac{H}{f_{\pi}}\text{ .} \label{masses}%
\end{equation}
It is clear that $m_{2}\rightarrow0$ for $H\rightarrow0$, i.e., this particle
represents the Goldstone boson emerging from the spontaneous 
breaking of chiral symmetry.

We now introduce polar coordinates $(\sigma,\pi)$ through the transformation%
\begin{equation}
\phi_{1}=\sigma\cos\frac{\pi}{\phi}\text{ },\text{ }\phi_{2}=\sigma\sin
\frac{\pi}{\phi}\text{ ,}\label{polartransf}%
\end{equation}
leading to%
\begin{equation}
\mathcal{L}_{pol}=\frac{1}{2}\partial_{\mu}\sigma\partial^{\mu}\sigma+\frac
{1}{2}\frac{\sigma^{2}}{\phi^{2}}\partial_{\mu}\pi\partial^{\mu}\pi
+\frac{m^{2}}{2}\sigma^{2}-\frac{\lambda}{2}\sigma^{4}+H\sigma\cos{\frac{\pi
}{\phi}}\ \text{.}\label{lpol}%
\end{equation}
Just as above, one shifts the field $\sigma$ as $\sigma\rightarrow\sigma+\phi
$. At zero temperature the minimization of the potential leads to the same Eq.\
(\ref{fix1}) for the minimum $\phi=\varphi$. Also the zero-temperature
tree-level masses $m_{\sigma}$ and $m_{\pi}$ coincide with the
expressions of Eq.\ (\ref{masses}):%
\begin{equation}
m_{\sigma}^{2}=m_{1}^{2}=-m^{2}+6\lambda f_{\pi}^{2}\text{  , }m_{\pi}%
^{2}=m_{2}^{2}=\frac{H}{f_{\pi}}\text{ .}\label{massespolar}%
\end{equation}
\noindent
In order to extract $m_{\pi}^2$, we have expanded the cosine in Eq.\ (\ref{lpol}).

\subsection{Mathematical issues using polar coordinates: The Jacobian and the
integration intervals}

\label{nuts}

Denoting $\pi/\phi\equiv\theta$ and taking into account that the
Jacobian of the transformation in Eq.\ (\ref{polartransf}) is $\sigma$, the
partition function can be rewritten as
\begin{equation}
Z=\oint\limits_{-\infty}^{\infty}\mathcal{D}\phi_{1}\mathcal{D}\phi
_{2}\, \exp\left(-\int\limits_{0}^{1/T}d\tau\int_{V}d\vec{x}\,
\mathcal{L}_{E,cart}%
\right) =\oint\limits_{0}^{\infty}\mathcal{D}\sigma\,\sigma\oint\limits_{0}^{2\pi
}\mathcal{D}\theta\, \exp \left(-\int\limits_{0}^{1/T}d\tau\int_{V}d\vec{x}%
\,\mathcal{L}_{E,pol} \right)\text{ ,}\label{z}%
\end{equation}
\noindent where periodic boundary conditions are understood: $\phi_{1}%
(\tau,\vec{x})=\phi_{1}(\tau+1/T,\vec{x})$, etc. The suffix $E$ means that the
Lagrangians are considered in Euclidean space.

The r.h.s.\ of Eq.\ (\ref{z}) describes the partition function in polar
coordinates. Due to the fact that the Jacobian is not unity and that both
fields $\sigma$ and $\theta$ do not vary between $(-\infty,\infty),$ the
question arises if we can apply the usual Feynman rules to the Lagrangian
$\mathcal{L}_{pol}$.

One can rewrite the contribution of the Jacobian as
\begin{equation} \label{Jacobian}
\oint\limits_{0}^{\infty}\mathcal{D}\sigma\,\sigma\longrightarrow\oint
\limits_{0}^{\infty}\mathcal{D}(\sigma \phi) \, \exp \left[-\int\limits_{0}
^{1/T}d\tau\int_{V}d\vec{x}\,\left(-\Lambda_{\tau}\Lambda_x^{3} 
\ln\frac{\sigma}{\phi} \right) \right]\;,
\end{equation}
where $ \Lambda_{\tau} \Lambda_x^3$ is an
infinite constant of dimension energy$^4$. (In discretized Euclidean space-time
$\Lambda_\tau^{-1} =a_\tau$ is the lattice spacing in time and
$\Lambda_x^{-1}=a_x$ the lattice spacing in spatial direction.)
In the following, we also use the notations
\begin{gather}
\label{velt}
 \Lambda_{\tau} \Lambda_x^3 \equiv I \equiv \delta^4(0) 
\equiv T \sum_{n=-\infty}^{\infty} \int \frac{d^3 \vec{k}}{(2 \pi)^3}\; 1\;.
\end{gather}
It is evident that the term in the exponent of Eq.\ (\ref{Jacobian})
induces a divergent contribution to the 
effective action requiring
regularization. In the framework of dimensional regularization the
contribution of the Jacobian vanishes in virtue of Veltman's
rule \cite{Leibbrandt:1975dj}, see also the explicit perturbative analyses
performed in Refs.\
\cite{Chervyakov:2000,Argyres:2009em,Argyres:2009sw}. 
Since the transformation to polar coordinates turns
a renormalizable Lagrangian into a non-renormalizable one, one expects
a perturbative cancellation of divergent contributions from the 
non-renormalizable term $\frac{1}{2}\frac{\sigma^{2}}{\phi^{2}}
\partial_{\mu}\pi\partial^{\mu}\pi $ appearing in Eq.\ (\ref{lpol}) with
the divergent contributions from the Jacobian. Indeed, it was shown in
Ref.\ \cite{Argyres:2009sw} that divergent contributions, $\sim I$, 
arising from the momentum-dependent interaction term, exactly cancel
the vertices from the Jacobian order by order in a perturbative loop
expansion. Using a power-counting argument, 
we demonstrate explicitly in Appendix \ref{cancel} how this cancellation
works for the CJT effective potential.
However, it turns out that the cancellation does not occur for a
truncation of the CJT effective potential at a given loop order (e.g.\ in Hartree 
approximation where only double-bubble diagrams are included), but
only when higher-order loop contributions are taken into account
(see Appendix \ref{cancel}). We
nevertheless omit contributions $\sim I$, assuming that the aforementioned
cancellation has happened before studying a particular truncation
of the effective potential.
Therefore, we neglect the Jacobian from
the beginning, independent of the renormalization scheme,
and simultaneously omit terms $\sim I$ arising from the
momentum-dependent vertices
[in our study this only concerns the first term appearing in Eq.\
(\ref{stp})]. 

Apart from this general argument why to omit the Jacobian, 
we explicitly verified that the relevant features of our results are 
the same independent of the renormalization scheme (trivial
regularization, counter-term regularization, or dimensional
regularization scheme).
In addition, in Sec.\ \ref{elina2} we introduce
polar coordinates in a slightly modified manner which corresponds to a unit
Jacobian. Again, our conclusions remain unchanged.

We now turn to the extension of the range of integration over the 
fields $\sigma$ and $\pi$. 
The possibility to extend the angular integration interval,
$\int\limits_{0}^{2\pi}\mathcal{D}\theta\rightarrow\int\limits_{-\infty
}^{\infty}\mathcal{D}\theta$, originates from the periodicity of the integrand.
This point is subtle since it is in general not possible to split a path
integral over the interval $I=I_{1}\cup I_{2}$ into the sum of two path
integrals over the intervals $I_{1}$ and $I_{2}$, respectively. For potentials
of the form $U(\sigma)$ and for $2\pi$-periodic potentials $U(\sigma,\theta)$
(as in the present case) one can show that extending the range of integration
simply yields a countably infinite overall constant which can be absorbed into
a normalization constant \cite{Grahl:2009}.
The extension in the $\sigma$ direction from $(0,\infty)$ to $(-\infty
,\infty)$ can be achieved with the help of a modified Heaviside step function
defined in such a way that its contribution vanishes in dimensional
regularization \cite{Grahl:2009}.  

Besides the divergences $\sim I$ which cancel order by order in
perturbation theory, one encounters the standard UV divergences in
loop integrals, see Appendix \ref{app1}. In many-body resummation
schemes, the cancellation of these divergences is subtle and
commonly requires additional counter terms compared to those
encountered in perturbation theory \cite{vanHees:2002bv}.
Unfortunately, we were not able to identify these counter terms in the
polar coordinate representation, but we shall assume that they exist
and cancel the above mentioned standard divergences.

\subsection{Shift of the potential}

\label{shifting}

In this section we show how to circumvent the problems arising from the fact
that polar coordinates are ill-defined at the origin. Inspired by Ref.\
\cite{Argyres:2009em}, we first shift the potential along the $\phi_{1}$-axis
by an arbitrary amount $v>0$. In this way the global minimum realized at the
critical temperature (and above) is not located at $\phi=0$, but at $\phi=v$. 
After the shift, the Lagrangian (\ref{lcart}) reads

\begin{equation}
\mathcal{L}_{v}=\frac{1}{2}\partial_{\mu}\phi_{1}\partial^{\mu}\phi_{1}%
+\frac{1}{2}\partial_{\mu}\phi_{2}\partial^{\mu}\phi_{2}+\frac{m^{2}}%
{2}\left(  \phi_{1}-v\right)  ^{2}+\frac{m^{2}}{2}\phi_{2}^{2}-\frac{\lambda
}{2}\left[  \left(  \phi_{1}-v\right)  ^{2}+\phi_{2}^{2}\right]  ^{2}%
+H(\phi_{1}-v)\ .
\end{equation}
When performing the transformation to polar coordinates, Eq.\
(\ref{polartransf}), we obtain
\begin{gather}
\mathcal{L}_{v}=\mathcal{L}_{pol}-m^{2}v\,\sigma\,\cos{\theta}+\frac{m^{2}}%
{2}v^{2}
-\frac{1}{2} \lambda  v (v-2 \sigma  \cos{\theta} ) \left(2 \sigma ^2+v^2-2 \sigma  v \cos{\theta} \right)
 - H\, v  
\; . \label{lshift}%
\end{gather}
Note that obviously $\mathcal{L}_{v=0}=\mathcal{L}_{pol}$ from Eq.\
(\ref{lpol}), thus the study of Sec.\ \ref{sec1} can be regarded as a
special case of this more general treatment.
After performing a Taylor expansion of the trigonometric functions about
$\theta=0$ and shifting $\sigma\rightarrow\sigma+\phi$, we can easily
determine the additional tree-level contributions to the masses and the
interaction vertices.

\section{Results at nonzero $T$}

\label{res}

In this section we present the results at nonzero temperature for the model
described by the Lagrangian $\mathcal{L}_{v}$ for different cases.

For reasons of simplicity, numerical results in Secs.\ \ref{move1}
-- \ref{move2} were obtained using the so-called trivial
regularization: the vacuum part of the integrals is simply set
to zero, for details see Appendix \ref{app1}. For a discussion of the
alternative counter-term regularization and dimensional regularization schemes we refer to Sec.\ \ref{elina1}.

\subsection{$H\neq0$, $v=0$}
\label{move1}

In this case the system is described by the Lagrangian (\ref{lpol}) in Sec.\
\ref{sec1}. The effective potential in the CJT formalism reads
\begin{equation}
V_{\rm eff}=U\left(  \phi\right)  +\frac{1}{2}\sum_{i=\sigma,\pi}\int
\limits_{k} \left[\ln  G_{i}^{-1}\left(  k\right)+
 D_{i}^{-1}\left(k,  \phi\right)
\ G_{i}\left(  k\right)  -1\right]  +V_{2}\left[  \phi,G_\sigma,G_\pi
\right]  \ , \label{veffcjt}%
\end{equation}
where $U=-m^2 \phi^2/2 + \lambda \phi^4/2 -H \phi$ 
denotes the classical potential and the inverse tree-level
propagators read
\begin{equation}
D_{\sigma}^{-1}(k,\phi)=-k^{2}+m_{\sigma}^{2}=-k^{2}-m^{2}+6\lambda\phi
^{2}\ ,\text{ }D_{\pi}^{-1}(k,\phi)=-k^{2}+\frac{H}{\phi}\ .
\end{equation}
At nonzero $T$ the condensate and the masses become $T$-dependent functions
\begin{equation}
\varphi\rightarrow\varphi(T)\text{, }m_{\sigma}\rightarrow M_{\sigma
}(T),\text{ }m_{\pi}\rightarrow M_{\pi}(T)\text{ ,}%
\end{equation}
and the dressed propagators are given by
\begin{equation}
G_{\sigma}^{-1}(k)\equiv-k^{2}+M_{\sigma}^{2}\ \text{, }G_{\pi}^{-1}(k)\equiv
-Z^{2}k^{2}+M_{\pi}^{2}\ , \label{dressprop}
\end{equation}
\noindent where $Z$ is a wave-function renormalization factor for the pion. 

The term $V_{2}$ in Eq.\ (\ref{veffcjt}) is the contribution of
all 2PI vacuum graphs, $\phi$ denotes the connected 1-point function in
the presence of a source, and $G$ denotes the full connected 2-point 
function in the presence of the source. In general, $V_2$ consists of
infinitely many diagrams, which prohibits an explicit calculation of $V_2$.
 In practice, one therefore has to restrict oneself to
certain classes of diagrams. We shall use the so-called \textit{Hartree
approximation} where only double-bubble diagrams are taken into account:
\[
V_{2}=\frac{3}{2}\lambda\left[  \int\limits_{k}G_{\sigma}(k)\right]^{2}
-\frac{H}{8\phi^{3}}\left[  \int\limits_{k}G_{\pi}(k)\right]^{2}-\frac{1}
{2\phi^{2}}\left[  \int\limits_{q}G_{\sigma}(q)\right]  \left[  \int
\limits_{k}k^{2}G_{\pi}(k)\right]  \ .
\]

By extremizing the effective potential in Eq.\ (\ref{veffcjt}) we obtain the
following equations:
\begin{gather}
m^{2}\varphi-2\lambda\varphi^{3}+H=6\lambda\varphi\int\limits_{k}{G}_{\sigma
}(k)-\frac{H}{2\varphi^{2}}\int\limits_{k}{G}_{\pi}(k)+\frac{3H}{8\varphi^{4}%
}\left[  \int\limits_{k}{G}_{\pi}(k)\right]  ^{2}+\frac{1}{\varphi^{3}}\left[
\int\limits_{q}{G}_{\sigma}(q)\right]  \left[  \int\limits_{k}k^{2}G_{\pi
}(k)\right]  \ ,\label{firstservice}\\
M_{\sigma}^{2}=-m^{2}+6\lambda\varphi^{2}+6\lambda\int\limits_{k}{G}_{\sigma
}(k)-\frac{1}{\varphi^{2}}\int\limits_{k}k^{2}{G}_{\pi}(k)\ ,\label{nonlinH1}\\
M_{\pi}^{2}=\frac{H}{\varphi}-\frac{H}{2\varphi^{3}}\int\limits_{k}{G}_{\pi
}(k)\ ,\label{service}\\
Z^{2}=1+\frac{1}{\varphi^{2}}\int\limits_{k}{G}_{\sigma}(k)%
\ ,\label{secondservice}%
\end{gather}
where $\phi=\varphi$ denotes, in general, an extremum. At $T=0$ the
masses coincide with their tree-level values in Eq.\ (\ref{masses}) and
$\varphi=f_{\pi}$. Furthermore,
\begin{equation}
H=m_{\pi}^{2}\,f_\pi ,
\end{equation}%
\begin{equation}
\lambda=\frac{m_{\sigma}^{2}-m_{\pi}^{2}}{4f_\pi^{2}}\ ,\label{nonlinH2}%
\end{equation}%
\begin{equation}
m^{2}=\frac{m_{\sigma}^{2}-3m_{\pi}^{2}}{2}\ .\label{nonlinH3}%
\end{equation}
The following numerical values are used at zero temperature:
$m_{\sigma}=600$ MeV, $m_{\pi}=139.5$ MeV, and $f_{\pi}=92.4$ MeV.

\begin{figure}
[ptb]
\begin{center}
\includegraphics[
height=2.6237in,
width=6.357in
]%
{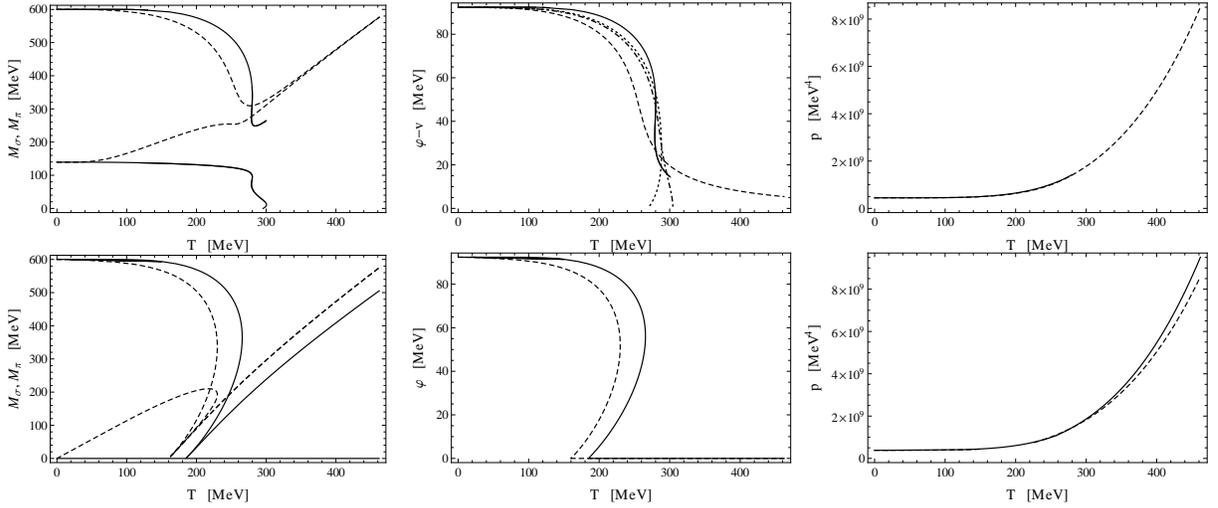}%
\caption{Upper row ($H\neq0$), from left to right: masses, extrema of the
effective potential, pressure (dashed lines: Cartesian, solid lines: polar, $v=0$).
Lower row ($H=0$), from
left to right: masses, extrema of the effective potential, pressure (dashed lines:
Cartesian, solid lines: polar, $v=0$). The dotted and dash-dotted 
lines in the middle panel of the upper row
correspond to the condensate in the polar case with $v=0.5f_{\pi}$ 
and $v=0.7325f_{\pi}$, respectively.}%
\label{paperfig1}%
\end{center}
\end{figure}

The numerical results for the masses, the condensate, and pressure can
be found in the upper row of Fig.\ \ref{paperfig1}. 
The results for polar coordinates are also
compared to the standard results obtained in Cartesian coordinates.
One observes that in general the results in polar coordinates differ
substantially from those in Cartesian coordinates, 
except for the pressure at low temperatures, where the agreement
between solid and dashed curves is seen to be quite good.
Moreover, although the condensate decreases sharply at a temperature
$\sim 300$ MeV, indicating the onset of chiral symmetry restoration,
the pion mass does not become degenerate with
the $\sigma$ mass for high $T$. On the contrary, above a temperature 
$T_{max}\simeq300.5$ MeV (where the curves in Fig.\ \ref{paperfig1} 
terminate) it becomes tachyonic,
signalizing a \emph{breakdown} of the model for large $T$. 
This fact shows a limitation of the
application of angular variables at nonzero temperature, at least
in the Hartree approximation. 

\begin{figure}[ptbh]
\includegraphics[scale=0.8]{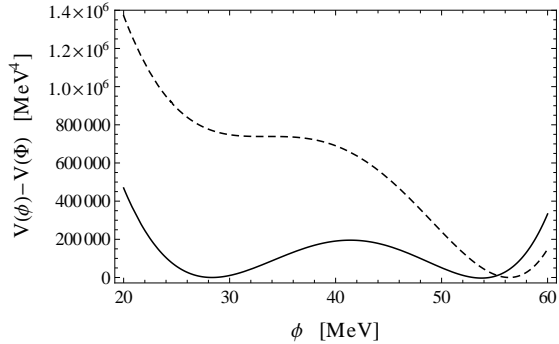}\caption{$H\neq0$, $v=0$.
Effective potential minus its value at its global minimum $\phi=\Phi$ for
a temperature $T=278.91$ MeV (dashed line) and $T^{\ast}=279.6$ MeV
(solid line).}%
\label{PAPERFIG2_1}%
\end{figure}

The effective potential is shown in Fig.\
\ref{PAPERFIG2_1} for different temperatures. 
There is a region around the origin where the effective
potential is not defined since no real-valued solutions exist. Due to the
singular terms with inverse powers of $\phi$ there is no
extremum at the origin. At low temperature there is only a global minimum at
a large value $ \phi \equiv \Phi>0$. 
At a certain temperature a second minimum and a maximum (both at
smaller values of $\phi$) occur. At
$T^{\ast}\simeq279.6$ MeV the effective potential assumes the
same value at both minima, indicating 
a first-order phase transition. Above this 
temperature, the global minimum $\Phi$ moves closer and closer to the
origin, but never becomes zero.
Above $T_{max}\simeq300.5$ MeV no real
solutions to the system of equations exist at the global minimum.

\begin{figure}[ptbh]
\includegraphics[scale=0.54]{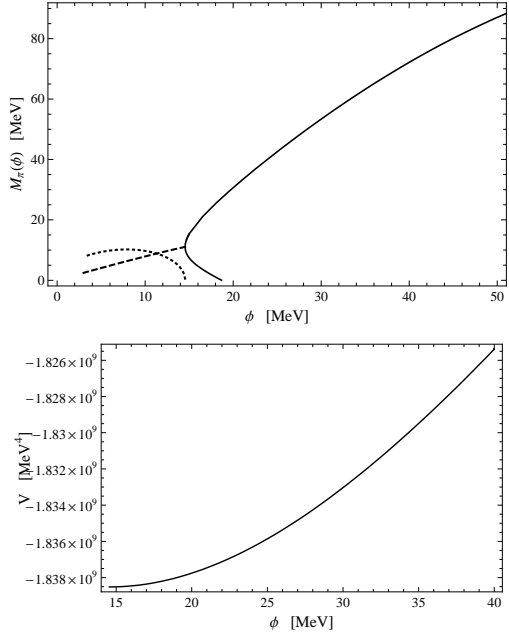}\caption{$H\neq0$, $v=0$, $T=300.5$
MeV $\simeq T_{max}$. Upper figure: Pion mass as function of the variable
$\phi.$ Solid line: $M_{\pi}(\phi)$ when only the real solution exists,
$\phi\geq \phi_{turn}$. Dashed line: Re$M_{\pi}(\phi)$, dotted line:
Im$M_{\pi}(\phi)$. Lower figure: effective potential as function of the
variable $\phi.$ The latter can be plotted only for $\phi\geq \phi_{turn}$, i.e.,
where real-valued solutions for the pion mass exist. For $\phi< \phi_{turn}$
also the effective potential becomes a complex function.}%
\label{PAPERFIG4}%
\end{figure}

The reason why no real solution exists above $T_{\max}$ is due to the fact
that the pion mass becomes imaginary, and therefore the system is not stable.
In order to see this in more detail, we show the {\it function\/}
$M_{\pi}(\phi)$ at $T=T_{max}$ in Fig.\ \ref{PAPERFIG4} (solid line).
We observe that $M_\pi(\phi)$ has an imaginary part below
a value $\phi_{turn} \simeq 15$ MeV. At $\phi_{turn}$ the imaginary part vanishes
and $M_\pi$ becomes a positive definite, real- (but multi-)valued function of
$\phi$. At $T=T_{max}$, 
the value $M_{\pi}(\phi =\phi_{turn})$
coincides with $M_\pi$ at the global minimum of the effective potential. 
The solid line above $\phi_{turn}$ is the solution for $M_{\pi}(\phi)$ 
which has to be used to plot the effective
potential. Using the lower branch instead would lead to a
discontinuous jump of the
effective potential near $\phi= 19$ MeV. 
Below $T_{max}$ the point $\phi_{turn}$
is located such that
the extremum $\varphi>\phi_{turn}$, i.e., real solutions at the 
global minimum $\Phi$
exist. With increasing temperature $\Phi$ approaches $\phi_{turn}$ and
finally hits the ill-defined region for $T_{max}$. 



\subsection{$H=0$, $v=0$}

In this section we study the case where the explicit
symmetry breaking parameter is set to zero,
$H = 0$. 
Equations (\ref{firstservice})--(\ref{secondservice}) simplify to
\begin{gather}
m^{2}\varphi-2\lambda\varphi^{3}=6\lambda\varphi\int\limits_{k}{G}_{\sigma
}(k)\ ,\text{ }M_{\sigma}^{2}=-m^{2}+6\lambda\varphi^{2}+6\lambda\int
\limits_{k}{G}_{\sigma}(k)\;, \label{mussichdenn}\\
Z^{2}=1+\frac{1}{\varphi^{2}}\int\limits_{k}{G}_{\sigma}(k)\ ,\text{ }M_{\pi}%
^{2}=0\; . \label{jetztnoch}%
\end{gather}
Here, we have used Eq.\ (\ref{A5}) from Appendix \ref{app1} and the
fact that $M_\pi^2 = 0$ to eliminate the pion tadpole term
in Eq.\ (\ref{service}).
The numerical results for the masses, the condensate, and the pressure
as functions of temperature are shown in the lower row of 
Fig.\ \ref{paperfig1}.
Note that $M_{\pi}^{2}=0$ indicates that the Goldstone theorem
is fulfilled at each $T$ below the chiral phase transition. 
This is a property which
does not hold in Cartesian coordinates, see for instance the
lower left panel of Fig.\ \ref{paperfig1}. 
Unfortunately, $M_\pi^2=0$ also above the transition, indicating that
the chiral partners do not become degenerate in the chirally restored
phase where $\varphi=0$.

\begin{figure}[ptb]
\begin{center}
\includegraphics[
height=2.1162in,
width=3.3797in
]
{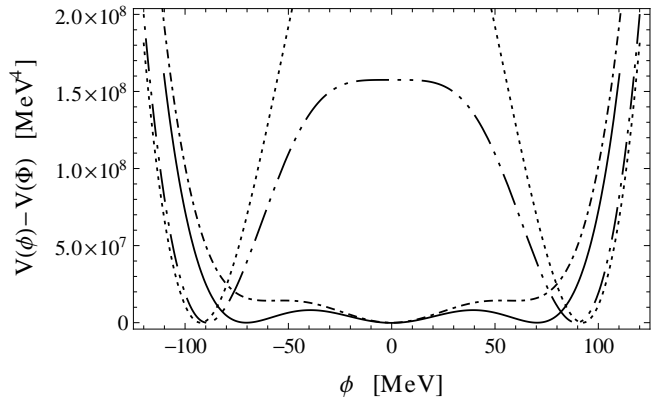}
\end{center}
\caption{$H=0$, $v=0$. Effective potential minus its value at its
global minimum for $T=0$ (dotted line), $T=1.998f_{\pi}$ (dot-dot-dashed line), $T^{\ast
}=2.787f_{\pi}$ (solid line) and $T=2.87f_{\pi}$ (dot-dashed line).}%
\label{paperfig2}%
\end{figure}

The effective potential (relative to its value at its
global minimum) for different temperatures
is shown in Fig.\ \ref{paperfig2}. For temperatures 
in the transition region, one clearly
observes the features of a first-order phase transition, i.e., 
three minima separated by two maxima.

\subsection{$H\rightarrow 0$, $v=0$}

\label{chirallimit}

The chiral limit, i.e., $H \rightarrow 0$, does not exist, since
for arbitrarily small, but nonzero values of $H$, the effective
potential has no extremum near the origin $\varphi=0$.
We have confirmed this
numerically by taking successively smaller values of $H$ 
(compare discussion of Fig.\ \ref{polarlim}).
The reason for this non-analyticity of the chiral limit are the
inverse powers of $\varphi$ in Eqs.\ 
(\ref{firstservice})--(\ref{secondservice}). When setting 
$H=0$ from the beginning, this problem does not exist because the
singular terms are eliminated right away.
Therefore, we conclude that the model cannot
describe the chirally restored phase and is limited to the
chirally broken phase.
Again, polar coordinates have demonstrated a limited range of
applicability. Interestingly, this problem does not
exist in Cartesian coordinates where taking the limit $H\rightarrow
0$ is not problematic in the chirally restored phase.

\subsection{$H\neq0$, $v\neq0$}
\label{move2}

As argued in Refs.\  \cite{Argyres:2009em,Argyres:2009sw},
the non-analyticity at the origin $\varphi = 0$
encountered in the previous subsections
can be eliminated by introducing a non-vanishing value for
the parameter $v$, i.e., shifting the vacuum expectation value
to nonzero values $\varphi = v$. Now we have to
consider the full Lagrangian (\ref{lshift}). In the double-bubble
approximation, the following equations are obtained:
\begin{gather}
0 = -m^2 (\varphi-v) + 2 \lambda (\varphi-v)^3 - H + 6 \lambda(
\varphi -  v ) \int\limits_{k} G_{\sigma}(k)    \nonumber \\
+ \left[ \frac{m^2 v-H}{2\varphi^2} + \lambda (\varphi-v) \left(\frac{v}{\varphi} + \frac{v^2}{\varphi^2} \right) \right]
\int\limits_{k} G_{\pi}(k) -\left(\frac{3}{8} \frac{  m^2 v-H -2v^3
    \lambda}{\varphi^4} 
+ \frac{4v^2 \lambda}{\varphi^3} - \frac{v \lambda}{4\varphi^2} 
\right) \left[\int\limits_{k} G_{\pi}(k)\right]^2    \nonumber \\
+\frac{1}{\varphi^3} \left[\int\limits_{q} G_{\sigma}(q)\right] 
\left[\int\limits_{k} k^2 G_{\pi}(k)\right] 
- \left(-\frac{4v^2 \lambda}{\varphi^3} + \frac{3v \lambda}{\varphi^2}
\right) 
\left[\int\limits_{q} G_{\sigma}(q)\right] \left[\int\limits_{k}
  G_{\pi}(k)
\right]  \ ,  \label{sys1}   
\end{gather}
\begin{gather}
M_{\sigma}^2 = 6 \lambda (\varphi-v)^2 - m^2  
+ 6 \lambda \int\limits_{k} G_{\sigma}(k) 
- \frac{1}{\varphi^2} \int\limits_{k} k^2 G_{\pi}(k) + 
2 \lambda \,\frac{v}{\varphi} \left( 3 - 2\,\frac{v}{\varphi} 
\right) \int\limits_{k} G_{\pi}(k)   \ ,  
\end{gather}
\begin{gather}
M_{\pi}^2 = \frac{H-m^2 v}{\varphi} 
  +2 \lambda \frac{v}{\varphi} (\varphi-v)^2
 - 2\lambda \frac{v}{\varphi} \left(2 \frac{v}{\varphi} -3 \right) 
\int\limits_{k} G_{\sigma}(k)   \nonumber \\  
-\left[\frac{H-m^2 v}{2 \varphi^3} +\lambda \frac{v}{\varphi^3} (\varphi-v)^2 - 6 \lambda \frac{v^2}{\varphi^2}  
  \right] \int\limits_{k} G_{\pi}(k)  \ ,    
\end{gather}
\begin{gather}
Z^2 = 1 + \frac{1}{\varphi^2} \int\limits_{k} G_{\sigma}(k) \ . \label{sys4}
\end{gather}

In the limit $v\rightarrow\infty$ the system of Eqs.\ 
(\ref{sys1})--(\ref{sys4}) reduces to the system of equations 
for Cartesian coordinates [see
 Eqs.\ (28a,b), (30a) in Ref.\ \cite{Lenaghan:1999si} for $N=2$ with the
replacement $\varphi \rightarrow \varphi-v$].
This is easily explained by the fact that the radial
as well as the angular variable are defined relative to the origin so that for
large $\phi$ we have $\sin{\frac{\pi}{\phi}}\simeq\frac{\pi}{\phi}$ and
$\cos{\frac{\pi}{\phi}}\simeq1$. Hence, in the limit $v\rightarrow\infty$
polar coordinates become Cartesian:
\[
\phi_{1}=\phi\cos{\frac{\pi}{\phi}}+\sigma\cos{\frac{\pi}{\phi}}\simeq
\phi+\sigma\ ,\text{ }\phi_{2}=\phi\sin{\frac{\pi}{\phi}}+\sigma\sin{\frac
{\pi}{\phi}}\simeq\pi\; .
\]

Solving the system of Eqs.\ (\ref{sys1})--(\ref{sys4}) in the case $H\neq0$
yields the condensate $\varphi-v$ shown in the middle panel of the
upper row of Fig.\
\ref{paperfig1} (dotted and dash-dotted lines). 
For values $v \lesssim 0.73f_{\pi}$ (for instance $v=0.5f_\pi$, dotted line)
the manner in which the condensate
$\varphi(T)$ is multi-valued reminds of the behavior for a first-order phase
transition.
However, for this to be the case, there would have to be
a third solution for $\varphi-v$ near the origin. 
This solution does not exist and therefore there is neither a
first-order transition nor a restored phase. 
Above $v \gtrsim 0.73 f_{\pi}$ (dash-dotted line)
the behavior of $\varphi(T)$ smoothly 
approaches the known Cartesian behavior, as expected.

\subsection{The limit $m_{\sigma}\rightarrow\infty$}

In this section we study the properties of the nonlinear
sigma model \cite{Bochkarev:1995gi} in polar coordinates at nonzero
temperature. 
This investigation
is also interesting in comparison to chiral perturbation
theory at nonzero temperature, see also Ref.\ \cite{Leupold:2004zj}
and refs.\ therein.

The nonlinear sigma model is obtained by taking the limit $m_{\sigma
}\rightarrow\infty$ and keeping the ratio $\lambda/m^{2}$ fixed. 
Inserting Eqs.\ (\ref{nonlinH2}) and (\ref{nonlinH3}) into Eq.\ 
(\ref{nonlinH1}) we obtain
\begin{equation}
M_{\sigma}^{2}= \frac{3m_{\pi}^{2}}{2f_\pi^2} 
\left[ f_\pi^2-\varphi^{2} - \int_k G_\sigma(k) \right]
-\frac{1}{\varphi^{2}}\int\limits_{k}k^{2}{G}_{\pi}(k)  
+\frac{m_{\sigma}^{2}}{2 f_\pi^2} \left[ 3 \varphi^{2} - f_\pi^2 
+ 3\int\limits_{k}{G}_{\sigma}(k)\right]
\ .\label{true}%
\end{equation}
For $m_\sigma \rightarrow \infty$, the last term dominates. Therefore,
if this term is positive definite, also $M_\sigma^2 \rightarrow \infty$. 
We have to demand that the last term is positive definite
in order to have real-valued solutions for $M_\sigma$.
Since $\int_k G_\sigma(k) \rightarrow 0$ for $M_\sigma \rightarrow
\infty$, the positive-definiteness of the last term requires that
\begin{equation}
\varphi \geq \frac{f_{\pi}}{\sqrt{3}}.
\end{equation}
This means that the range $\varphi < f_\pi/\sqrt{3}$ cannot be described, 
neither for $H\neq0$ nor for $H\rightarrow 0$ (i.e.,
$m_{\pi}\rightarrow 0$). Thus, the model is not applicable
in the chirally restored phase where we expect $\varphi \simeq 0$.
This result shows that it
is not possible to describe the phase transition in polar
coordinates (in the framework of the CJT
formalism) when taking the limit to the nonlinear sigma model.

\subsection{Dependence on the regularization procedure}

\label{elina1}

It is important to verify how the results change when a different
regularization scheme is employed. In this section we compare three
different regularization schemes: the trivial regularization scheme, the
counter-term regularization scheme of Ref.\ \cite{Lenaghan:1999si},
and the dimensional regularization scheme of Ref.\
\cite{Andersen:2004fp}.
In the latter two cases we assume that suitable counter terms exist
for the Hartree approximation in polar
coordinates, see remark at the end of Sec.\ \ref{nuts}. 

In the counter-term scheme, 
the vacuum contribution of the thermal tadpole integral (\ref{G}),
\begin{equation}
\int\frac{d^{3}\vec{k}}{\left(  2\pi\right)  ^{3}}\frac{1}{2\sqrt{\vec{k}
^{2}+M^{2}}} \ ,
\end{equation}
is not neglected as in the trivial regularization scheme but is first
rewritten using the residue theorem,
\begin{equation}
\int\frac{d^{4}k}{\left(  2\pi\right)  ^{4}}\frac{1}{k^{2}+M^{2}}\text{ .}%
\end{equation}
This integral is then regularized by introducing a renormalization scale $\mu$
\cite{Lenaghan:1999si}:
\begin{equation}
Q_{\mu} (M)=\int\frac{d^{4}k}{\left(  2\pi\right)  ^{4}}\left[  \frac{1}%
{k^{2}+M^{2}}-\frac{1}{k^{2}+\mu^{2}}-\frac{\mu^{2}-M^{2}}{\left(  k^{2}%
+\mu^{2}\right)  ^{2}}\right]  =\text{ }\dfrac{1}{(4\pi)^{2}}\left[  M^{2}%
\ln\left(  \dfrac{M^{2}}{\mu^{2}}\right)  -M^{2}+\mu^{2}\right]  \text{ .}
\label{CT}%
\end{equation}
We set $\mu=m_{\sigma}$ for the counter-term
scheme and $\mu = m_{\sigma} / \sqrt{e}$ for the dimensional
regularization scheme, in order to satisfy the constraint 
$Z(T=0)=1$ for the vacuum value of the pion wave-function
renormalization factor.
The initial values $M_{\sigma} (T=0)= m_{\sigma}$, $M_{\pi} (T=0) = m_{\pi}$ 
and $\varphi (T=0) = f_{\pi}$ correspond to the following parameter choice:
\begin{gather}
m^2 =\frac{1}{2} \left\{m_{\sigma }^2  -m_{\pi }^2 
\left[\frac{5 Q(m_{\pi})}{4 f_{\pi }^2}+\frac{3}{2}+\frac{3 f_{\pi }^2}{2 f_{\pi
    }^2-Q(m_{\pi})}\right]\right\} \; , \\
\lambda = \frac{ 4f_\pi^2 [2f_{\pi}^2-Q(m_{\pi}) ](m_{\sigma}^2
  -m_{\pi}^2)  -m_{\pi}^2 Q^2(m_{\pi})  }{  16 f_{\pi}^4 
\left[2 f_{\pi}^2 - Q(m_{\pi})\right]} \; , \\
H = \frac{2 f_{\pi}^3 m_{\pi}^2}{2f_{\pi}^2 -Q(m_{\pi})} \; ,
\end{gather}
where $Q(m_{\pi}) \equiv Q_{\mu}(m_{\pi})$ from Eq.\ (\ref{CT})
for the counter-term scheme, 
and $Q(m_{\pi}) \equiv Q_{\mu}^{DR}(m_{\pi})$ from Eq.\ (\ref{QmuDR})
for the dimensional regularization scheme.


\begin{figure}[htbp]
\includegraphics[scale=0.375]{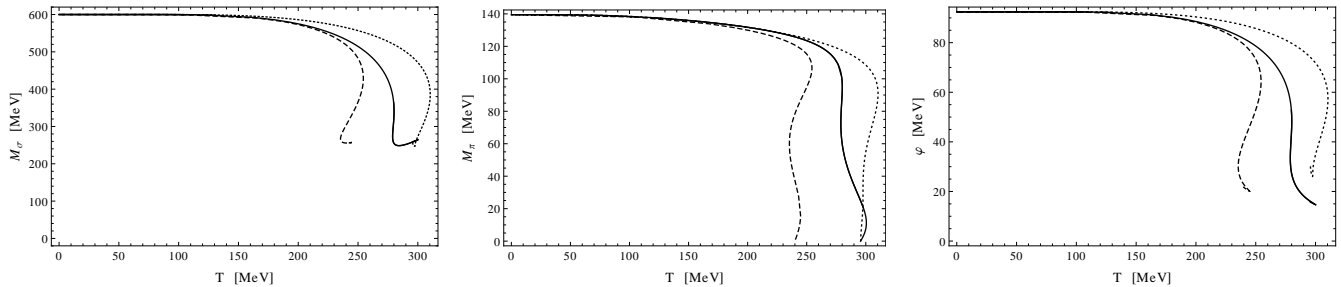}
\caption{The sigma mass (left panel), the pion mass (middle panel),
and the condensate (right panel) as functions of
temperature in the case $H \neq 0,\, v=0$ in the trivial
regularization (solid line), the dimensional regularization scheme
with $\mu=m_{\sigma} / \sqrt{e}$ (dotted line), and the
counter-term regularization scheme with $\mu=m_{\sigma}$ (dashed line).}
\label{polarexp}
\end{figure}

\begin{figure}[htbp]
\includegraphics[scale=0.38]{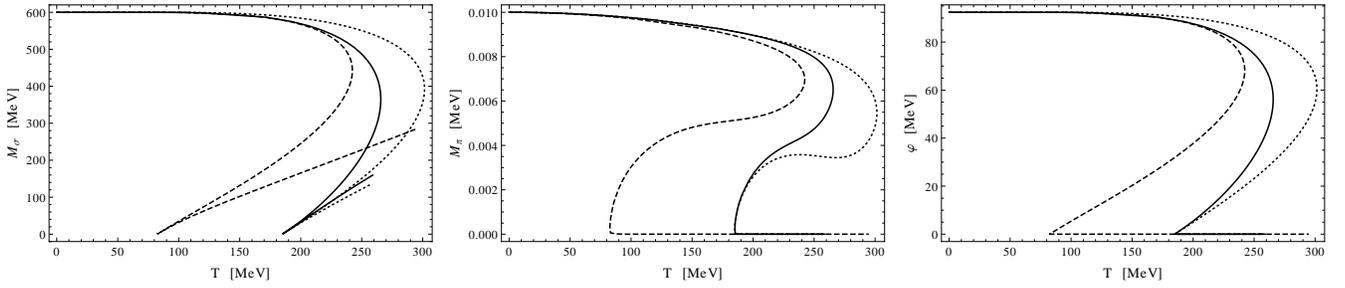}
\caption{The sigma mass (left panel), the pion mass (middle panel),
and the condensate (right panel) as functions of
temperature in the case $H \rightarrow 0,\, v=0$. Results for trivial
regularization are shown as solid lines, for 
dimensional regularization with $\mu=m_{\sigma} / \sqrt{e}$ as dotted
lines, and for counter-term regularization with $\mu=m_{\sigma}$ as dashed lines. 
For the middle panel, the vacuum pion mass was chosen to be
$m_{\pi}=10^{-2}$ MeV, but the shape of the curves remain
visually indistinguishable if we use smaller vacuum pion masses and
perform a rescaling of the $y-$axis. Only the value for the temperature 
where the pion mass becomes tachyonic (which happens where the curves end), $T_{end}$, changes.}
\label{polarlim}
\end{figure}

%

Figures \ref{polarexp} and \ref{polarlim} show the temperature
dependence of the condensate
and the masses applying the
trivial regularization, the dimensional regularization, and the
counter-term regularization
prescription, respectively.
As one can see, the choice for the regularization scheme yields qualitatively 
similar results. It is also not possible to avoid the pion
from becoming tachyonic at high temperatures, neither in the case with
explicitly broken symmetry, nor in the
chiral limit. In the latter case, cf.\ Fig.\ \ref{polarlim}, Goldstone's theorem 
is fulfilled since the pion is massless
in the phase of broken symmetry. Nevertheless, as with the trivial regularization
scheme, since the pion becomes tachyonic, no chirally restored phase
exists beyond a certain $T_{end}$ when approaching $H = 0$.


\section{The instructive example of the free Lagrangian}

\label{free}

In order to explain the problems related to polar coordinates, in this
section we consider a free Lagrangian in Cartesian coordinates:
\begin{equation} \label{Lfree}
\mathcal{L}_{free-cart}=\frac{1}{2}\partial_{\mu}\vec{\phi} \cdot 
\partial^{\mu}\vec{\phi}-\frac{m^{2}}{2}\,\vec{\phi} \cdot \vec{\phi}\text{ ,}%
\end{equation}
where $m^2 >0$. The pressure can be calculated exactly:
\[
p_{free-cart}=-\frac{1}{2}\int\limits_{k}\ln D{_{\sigma}^{-1}}-\frac{1}{2}%
\int\limits_{k}\ln D{_{\pi}^{-1}} \ , 
\]
with \noindent$D_{\sigma,\pi}^{-1}(k,\phi)=-k^{2}+m_{\sigma,\pi}^{2} 
= -k^2+m^2$.
Explicitly:%
\begin{equation}
p_{free-cart}=-\frac{T}{\pi^{2}}\int\limits_{0}^{\infty}dkk^{2}\ln\left(
1-e^{-\frac{\sqrt{k^{2}+m^{2}}}{T}}\right) \ .  \label{pfreecart}%
\end{equation}

Let us now perform the transformation to polar coordinates. The Lagrangian
\begin{equation}
\mathcal{L}_{polar}=\frac{1}{2}\partial_{\mu}\sigma\partial^{\mu}%
\sigma+\frac{1}{2}\frac{\sigma^{2}}{\phi^{2}}\partial_{\mu}\pi\partial^{\mu
}\pi-\frac{m^{2}}{2}\sigma^{2}
\end{equation}
\noindent
follows directly from Eq.\ (\ref{lpol}) by
setting $\lambda=0$ and replacing $m^{2}\rightarrow-m^{2}$. Note that,
although the original Cartesian Lagrangian (\ref{Lfree}) is that of a
noninteracting
system, the transformation to polar coordinates introduces a
momentum-dependent four-particle interaction.

The effective potential in double-bubble approximation reads
\begin{equation}
V_{eff}  =  \frac{m^2}{2}\, \phi^2 + \frac{1}{2} \int_k \left[ \ln
  G_\sigma^{-1}(k) + \ln G_\pi^{-1}(k) 
+ D_\sigma^{-1}(k) G_\sigma(k) + D_\pi^{-1}(k) G_\pi(k) -2
\right]  -\frac{1}{2\phi^2} \left[ \int_q G_\sigma(q) \right] \left[ \int_k k^2
  G_\pi(k) \right]\;.  \label{effpotfree}
\end{equation}
The (inverse) tree-level propagators are (as usual neglecting contributions $\sim
I$ from the Jacobian)
\begin{equation}
D_\sigma^{-1}(k) = -k^2 + m^2\;,\;\;\; D_\pi(k) = -k^2\;.
\end{equation}
The stationarity condition for $\phi$ is simply
\begin{equation}
0 = m^2 \varphi + \frac{1}{\varphi^3} \int_q G_\sigma(q) \int_k k^2 G_\pi(k)\;.
\end{equation}
Due to Eq.\ (\ref{A5}), the last integral is 
proportional to $M_\pi^2$. Therefore, if $M_\pi = 0$, the only
solution of the stationarity condition is $\varphi=0$, as one would expect
since no symmetry is broken.

The equations for the masses and the wave function renormalization
constant are given by 
\begin{eqnarray}
M_{\sigma}^{2}& =& m^{2} - \frac{1}{\varphi^2} \int_k k^2 G_\pi(k)\;,\\
M_\pi^2 & = & 0\;,\\
Z^{2}& =& 1 + \frac{1}{\varphi^2} \int_k G_\sigma(k)\;.
\end{eqnarray}
Because the tadpole $\int_k k^2 G_\pi(k)$ vanishes
for $M_\pi = 0$, see Eq.\ (\ref{A5}), we obtain the very simple set
of stationarity conditions
\begin{equation}
\varphi = 0\;,\;\;\; M_\sigma = m\;,\;\;\; M_\pi = 0\;, \;\;\; Z = \infty\;.
\end{equation}
Inserting this solution into the effective potential
(\ref{effpotfree}),
we obtain the pressure in polar coordinates (and in double-bubble
approximation)
\begin{equation}
p_{polar}\equiv - V_{eff}(\varphi) =
- \frac{1}{2} \int_k \left[ \ln D_\sigma^{-1}(k) + \ln D_\pi^{-1}(k)
+ 2 \ln Z  - 1 \right]\;.
\end{equation}
Ignoring the last two terms in brackets, we obtain
\begin{equation}
p_{polar} =
-\frac{T}{2\pi^{2}}\int\limits_{0}^{\infty}dkk^{2}\ln\left(
1-e^{-\frac{\sqrt{k^{2}+m^{2}}}{T}}\right)  +\frac{T^{4}\pi^{2}%
}{90}\ .
\end{equation}
The pressure $p_{polar}$ represents the sum of one
particle with mass $m$ and one particle with mass $m=0$. This latter
contribution is, however, not correct and overestimates the exact 
result of Eq.\ (\ref{pfreecart}). In principle, due to the $S$-matrix
equivalence
theorem, the result for the free theory should be the same in
Cartesian and
polar coordinates. The inequivalence found in the double-bubble
approximation demonstrates the inadequacy of this approximation when
polar coordinates are used. We suspect that an infinite resummation of
a certain class of (or of all) higher-loop 2PI diagrams is required in
order
to prove the equivalence between the Cartesian and the
polar-coordinate
representation. This simple example shows once more 
that care is needed
when polar coordinates (and a certain many-body approximation) 
are used to study properties of a system at nonzero
$T$.

\section{An alternative way to introduce polar coordinates}

\label{elina2}

\subsection{The transformation}


As we have discussed in Sec.\ II B, the Jacobian associated with the
transformation to polar coordinates introduced in Eq.\ (\ref{polartransf}) is
not unity. In this section we present an alternative polar representation 
$(\psi,\pi)$ for the $O(2)$ linear $\sigma$-model, which is defined as:
\begin{equation}
\phi_{1}=2\sqrt{\phi\psi}\cos\frac{\pi}{2\phi}\text{ , }\phi_{2}=2\sqrt
{\phi\psi}\sin\frac{\pi}{2\phi}\text{ .}\label{2polar}%
\end{equation}
In this case the associated Jacobian remains unity. Here the massive scalar
field acquiring a nonvanishing vacuum expectation value in the case of
spontaneously broken symmetry is represented by $\psi$.

The Lagrangian (\ref{lcart}) expressed in terms of the fields $\psi$ and $\pi$
introduced in Eq. (\ref{2polar}) reads
\begin{align}
\mathcal{L} &  =\dfrac{1}{2}(\partial_{\mu}\psi)^{2}\ +\dfrac{1}{2}%
(\partial_{\mu}\pi)^{2}-2m^{2}\phi^{2}-8\lambda\phi^{4}+2H\phi-\dfrac{\psi
^{2}}{2}\left(  16\lambda\phi^{2}+\dfrac{H}{2\phi}\right)  -\dfrac{\pi^{2}}%
{2}\dfrac{H}{2\phi}\nonumber\\
&  -(\partial_{\mu}\psi)^{2}\dfrac{\psi}{2\phi}+(\partial_{\mu}\pi)^{2}%
\dfrac{\psi}{2\phi}\ -\dfrac{H}{8\phi^{2}}\psi\pi^{2}+\dfrac{H}{8\phi^{2}}%
\psi^{3}+(\partial_{\mu}\psi)^{2}\dfrac{\psi^{2}}{2\phi^{2}}-\dfrac
{\ 5H}{64\phi^{3}}\psi^{4}+\dfrac{H}{\ 192\phi^{3}}\pi^{4}+\dfrac{\ H}%
{32\phi^{3}}\psi^{2}\pi^{2}\nonumber\\
&  +\text{higher-order terms.}\label{Lag4}%
\end{align}
where the shift $\psi\rightarrow\psi+\phi$ has also been performed. 

The corresponding inverse tree-level propagators and tree-level masses are
given by:
\begin{align}
D_{i}^{-1}(k;\phi)  &  =-k^{2}+m_{i}^{2},\text{ }i=\psi,\pi\nonumber\\
m_{\psi}^{2}  &  =16\lambda\phi^{2}+\dfrac{H}{2\phi},\text{ }m_{\pi}%
^{2}=\dfrac{H}{2\phi}\text{ }.
\end{align}

Applying the CJT formalism we derive the effective potential within the
double-bubble approximation:%
\begin{align}
\ V_{eff}(\phi;G_{\pi},G_{\psi})  &  =2m^{2}\phi^{2}+8\lambda\phi^{4}%
-2H\phi+\dfrac{1}{2}\int[\ln G_{\psi}^{-1}(k)+D_{\psi}^{-1}(k;\phi)G_{\psi
}(k)-1]\nonumber \\
&  +\dfrac{1}{2}\int[\ln G_{\pi}^{-1}(k)+D_{\pi}^{-1}(k;\phi)G_{\pi
}(k)-1]\ +V_{2}(\phi;G_{\pi},G_{\psi})\text{ ,}%
\end{align}%
\begin{align}
V_{2}(\phi;G_{\pi},G_{\psi})  &  =-\dfrac{1}{2\phi^{2}}\int_{k}k%
{{}^2}%
G_{\psi}(k)\int_{l}G_{\psi}(l)+\dfrac{15H}{64\phi^{3}}\left[  \int_{k}G_{\psi
}(k)\right]  ^{2}\nonumber\\
&  -\dfrac{H}{64\phi^{3}}\left[  \int_{k}G_{\pi}(k)\right]  ^{2}-\dfrac
{\ H}{32\phi^{3}}\int_{k}G_{\psi}(k)\int_{k}G_{\pi}(k)\text{ }.
\end{align}

Finally, the stationary conditions for the effective potential give the
following equations for the temperature-dependent masses, the condensate,
and the wave-function renormalization for the $\psi$-field:
\begin{align}
2H  &  =4m^{2}\varphi+32\lambda\varphi^{3}+\left(  16\lambda\varphi-\dfrac
{H}{4\varphi^{2}\ }\right)  \int_{k}{G}_{\psi}(k)-\dfrac{H}{4\varphi^{2}%
\ }\int_{k}{G}_{\pi}(k)+\dfrac{1}{\varphi^{3}}\int_{k}k^{2}{G}_{\psi}%
(k)\int_{l}{G}_{\psi}(l)\nonumber\\
&  -\dfrac{\ 45H}{64\varphi^{4}}\left[  \int_{k}{G}_{\psi}(k)\right]
^{2}+\dfrac{\ 3H}{64\varphi^{4}}\left[  \int_{k}{G}_{\pi}(k)\right]
^{2}+\dfrac{\ 3H}{32\varphi^{4}}\int_{k}{G}_{\psi}(k)\int_{k}{G}_{\pi
}(k)\text{ },
\end{align}%
\begin{equation}
M_{\psi}^{2}=16\lambda\varphi^{2}+\dfrac{H}{2\varphi}+\dfrac{\ H}%
{16\varphi^{3}}\left[  \ 15\int_{k}{G}_{\psi}(k)-\int_{k}{G}_{\pi}(k)\right]
-\dfrac{1}{\varphi^{2}}\int_{k}k%
{{}^2}%
{G}_{\psi}(k)\text{ },
\end{equation}%
\begin{equation}
M_{\pi}^{2}=\dfrac{H}{2\varphi}-\dfrac{\ H\ }{16\varphi^{3}}\left[  \int
_{k}{G}_{\psi}(k)+\int_{k}{G}_{\pi}(k)\right]  \text{ },\text{ }\ Z_{\psi}%
^{2}=1+\dfrac{1}{\varphi^{2}}\int_{k}{G}_{\psi}(k)\text{ }.
\end{equation}

\subsection{Results}

\begin{figure}
[ptb]
\begin{center}
\includegraphics[
height=2.1616in,
width=4.7542in
]%
{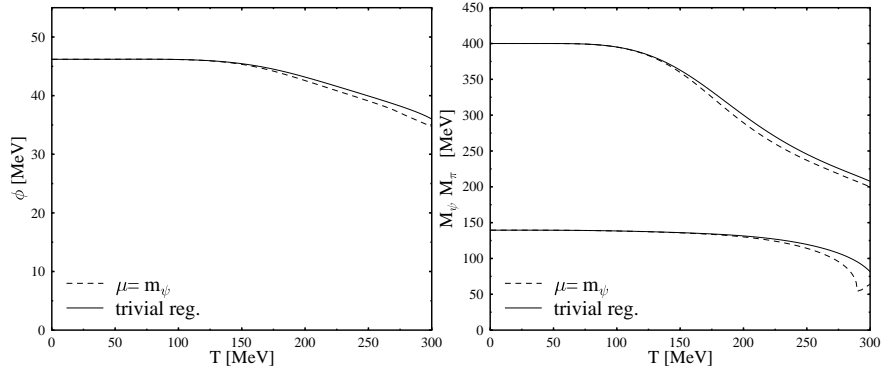}%
\caption{The temperature-dependent condensate (left panel) and
the meson masses (right panel) for the
alternative polar representation in the case of 
explicitly broken symmetry, in the trivial
regularization (solid lines), and in the 
counter-term regularization scheme (dashed lines) with $\mu=m_{\psi}.$
}%
\label{wurzelh}%
\end{center}
\end{figure}

\begin{figure}
[ptb]
\begin{center}
\includegraphics[
height=2.19in,
width=4.8172in
]%
{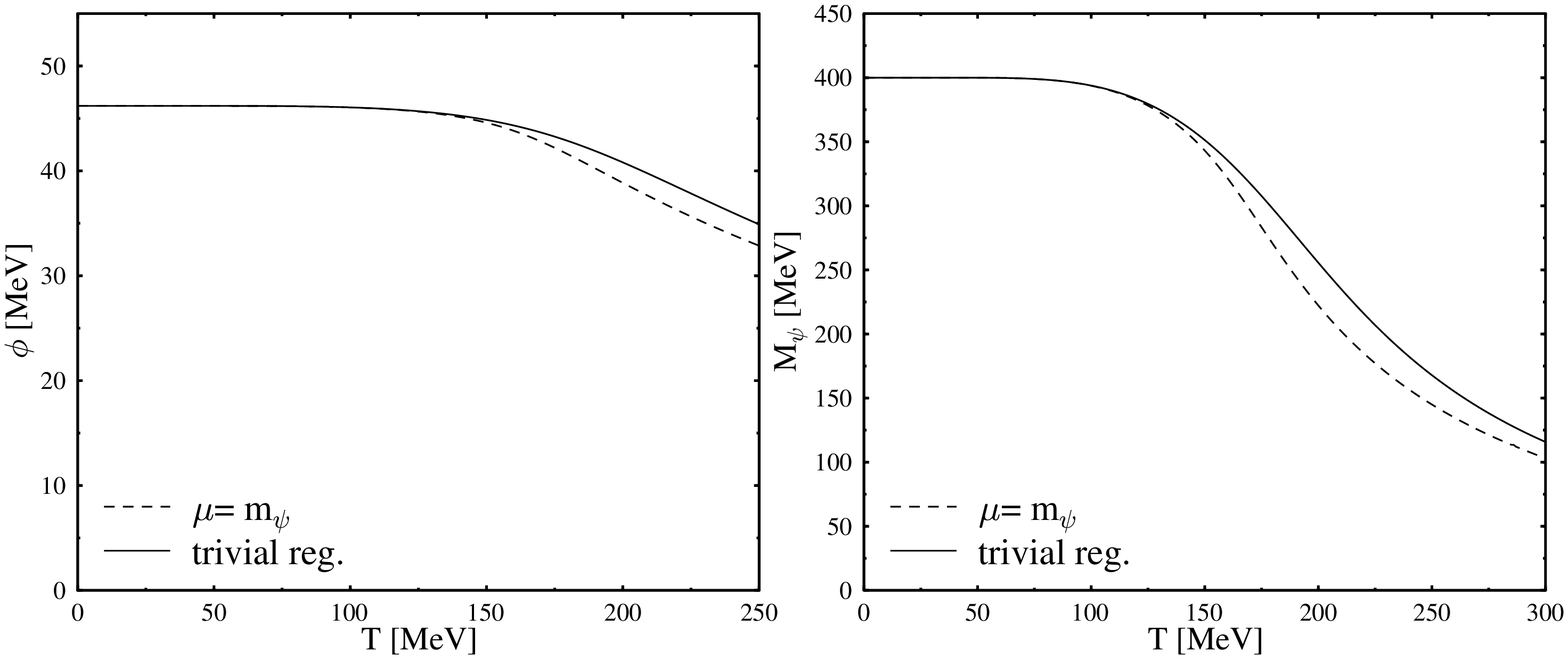}%
\caption{The temperature-dependent condensate and $\psi$-mass 
for the
alternative polar representation in the chiral limit, 
in the trivial regularization (solid line), and
in the counter-term regularization scheme (dashed line) with $\mu=m_{\psi}.$}%
\label{wurzelcl}%
\end{center}
\end{figure}

In this subsection we present the numerical results for this alternative polar
representation. Figure \ref{wurzelh} shows the condensate and the masses as
function of $T$ using the trivial and the counter-term regularization schemes
in the case of explicit chiral symmetry breaking. In Fig.\ \ref{wurzelcl} the
same quantities are shown in the chiral limit. The initial values $M_{\psi
}(T=0)=m_{\psi}$, $M_{\pi}(T=0)=m_{\pi},$ $Z_{\psi}(T=0)=1$, and $\varphi
(T=0)=f_{\pi}/2$ correspond to the following parameter choices. For trivial
regularization:
\begin{equation}%
\begin{array}
[c]{cccc}%
H\ =m_{\pi}^{2}f_{\pi}\ ,\text{ } &
m_{\ }^{2}\ =-\dfrac{m_{\psi}^{2}-3m_{\pi}^{2}}{2\ }\ ,\text{ } &
\lambda=\dfrac{m_{\psi}^{2}-m_{\pi}^{2}}{4f_{\pi}^{2}\ }\;,
\end{array}
\end{equation}
and for counter-term regularization:
\begin{gather}
H   =\dfrac{2f_{\pi}^3m_{\pi}^{2}}{2f_\pi^2-Q_{\mu}(m_{\pi})} \; , \;\;\;
\lambda  =\dfrac{1}{4f_{\pi}^{2}}\left( m_\psi^2-\dfrac{H}{2f_{\pi}^3}
\left[2f_\pi^2-Q_{\mu}(m_{\pi})\right]\right) \ , \ \
m^{2}   =\dfrac{H}{f_{\pi}} \left(1+\dfrac{4f_{\pi}^2Q_{\mu}(m_{\pi})-3Q_{\mu}^{2}(m_{\pi})}{8f_{\pi}^{4}} \right)
  -2\lambda f_{\pi}^{2}\ \ ,
\end{gather}
where we have to choose $\mu=m_{\psi}$ due to the condition $Z_{\psi}(T=0)=1.$

 In the
case of explicitly broken symmetry there is a crossover transition and
one observes no degeneration of the chiral partners at high $T$. Just as in
Sec.\ III, for $T\gtrsim300$ MeV the validity of the model breaks down and the
numerical results are no longer reliable. Moreover, chiral restoration is
approached very slowly, also when the chiral limit is taken.

Summarizing the results, the alternative polar representation does not offer a
better description of the chiral phase transition (at least not within
the chosen many-body approximation). The reason for this is
that, when constructing this alternative polar representation, one has to
evaluate several terms of the form $f(1+x)$ with $x=\psi/\phi$ and $x=\pi
/\phi$. Increasing the temperature the condensate starts melting while the
fluctuation of the fields become larger and the calculation becomes no longer
reliable for higher temperatures. Especially when approaching the critical
temperature all terms of the form $x/\phi,$ with $x=\pi,\psi$,
become problematic, limiting the validity of the model to lower temperatures.

\section{Conclusions}

\label{conclusionsoutlook}

In this paper, we have studied the 
$O(2)$ model in polar coordinates at
nonzero temperature. 
After having clarified some issues related to the transformation from
Cartesian to polar coordinates in the functional integral
representation of the partition function, we have computed the latter
in the CJT formalism in double-bubble approximation.
We have studied in detail the cases where the chiral symmetry is
explicitly broken and where the explicit symmetry breaking parameter
is set to zero. We have distinguished the latter case 
from the chiral limit, where the
explicit symmetry breaking parameter is smoothly sent to zero. We have
found that this limit does not exist in the strict mathematical sense,
due to non-analytic terms 
\[
\frac{1}{2}\frac{\sigma^{2}}{\phi^{2}}\partial_{\mu}\pi\partial^{\mu}%
\pi\ \ \mbox{and}\ \ H\sigma\cos{\frac{\pi}{\phi}} %
\]
\noindent 
appearing in the Lagrangian for polar coordinates.
Except when the explicit symmetry-breaking parameter
is exactly zero, we have found that, when approaching
$\phi\rightarrow 0$, the ensuing divergences are sufficiently
severe to invalidate the approach above a certain maximal
temperature $T_{max}$, above which no
physical solutions exist. 
The same results hold in the nonlinear limit $m_{\sigma}\rightarrow\infty.$

We have also investigated the possibility of shifting the potential by an
amount $v$ along the $\phi_{1}$-direction in order to circumvent the 
divergences resulting from the non-analytic terms in the Lagrangian.
Variation of $v$ allows to change smoothly from
polar coordinates to Cartesian coordinates (corresponding to $v\rightarrow
\infty$). Our conclusions about the suitability of polar coordinates
remain unchanged. However, $T_{max}$ increases with
$v,$ so that the range of applicability of polar coordinates is
extended. Moreover, the larger $v$, the better the agreement with the
Cartesian results. 

We have furthermore introduced and studied an alternative representation
for polar coordinates
(with unit Jacobian in the functional integral representation of the
partition function). Also in this case, the above mentioned problems
persist.

Our most important result is that, 
both in the chiral limit and in the case of explicitly
broken symmetry, the chiral partners do not become degenerate in mass
at high $T,$ not even when approaching $T_{max}$ where the order
parameter has already decreased by a substantial amont. In general,
the sigma particle becomes more massive while the pion mass decreases
or remains zero (in the absence of explicit symmetry breaking). Above
$T_{max}$ the pion even becomes tachyonic.
The absence of degeneracy of the chiral partners means that 
an important indication for the restoration of
chiral symmetry is missing when using polar coordinates and the
Hartree approximation.
We conclude that the use of angular variables is not well suited for 
the study of the chiral phase transition, at least in the Hartree
approximation.
 

A possible extension of the present study would be to use
 four-dimensional polar
coordinates, corresponding to the $O(4)$ model. The polar $O(4)$ model has
the advantage that three degrees of freedom can be identified with the three
pions, $\pi^{0}$ and $\pi^{\pm}$, and the remaining one with their chiral
partner, the sigma particle. The number of angular degrees of freedom could
affect the behavior of the equations in the limit $\varphi\rightarrow0$.
However, we believe that this generalization will not fix the problem
encountered in the polar version of the $O(2)$ model, since the divergences
pointed out above are general.

Finally, although on a conceptual level each representation is equivalent, on
a practical level the use of Cartesian coordinates is favorable to study
thermodynamical properties of systems described by the $O(N)$ model. In
connection to QCD, one should extend the model by incorporating all the
relevant low-energy mesons: besides scalar and pseudoscalar particles, also
vector and axial-vector degrees of freedom should be included in an enlarged
$U(N)\times U(N$) symmetry for a more realistic treatment of properties of QCD
at nonzero $T$ \cite{Gasiorowicz:1969kn,Ko:1994en,Struber:2007bm,Parganlija:2010fz}.

\bigskip

\bigskip

\textbf{Acknowledgment:} 
The authors thank Stefan Str\"uber, Hendrik van Hees, and Jochen Wambach 
for useful discussions. M.G.\ and E.S.\ thank HGS-HIRe for FAIR for funding.
D.H.R.\ thanks Kari J.\ Eskola and the department of physics 
of Jyv\"askyl\"a University for their warm hospitality during
a visit where part of this work was done.

\appendix

\section{Thermal integrals}

\label{app1}

In this appendix we list the standard thermal integrals which were used for numerical
calculations.  In our notation $k^2 = k_0^2-\vec{k}^2$ and 
\begin{gather}
 \int_k f(k_0,\vec{k}) \equiv T \sum_{n=-\infty}^{\infty} \int \frac{d^3 \vec{k}}{(2 \pi)^3} f(i 2\pi n T,\vec{k}) \ . 
\end{gather}
Carrying out the Matsubara summation of the thermal tadpole
integral gives
\begin{equation}
\int_{k} \frac{1}{-k^2+M^2} =\int\frac{d^{3}\vec{k}}{\left(  2\pi\right)  ^{3}}\frac{1}%
{\sqrt{\vec{k}^{2}+M^{2}}}\left[  \frac{1}{2}
+\frac{1}{\exp\left(  \sqrt{\vec{k}%
^{2}+M^{2}}/T\right)  -1}\right]  \text{ }, \label{G}%
\end{equation}
\noindent consisting of a finite contribution,
\[
Q_{T}(M)=\int\frac{d^{3}\vec{k}}{\left(  2\pi\right)  ^{3}}\frac{1}{\sqrt
{\vec{k}^{2}+M^{2}}}\frac{1}{e^{\sqrt{\vec{k}^{2}+M^{2}}/T}-1}=\int
\limits_{0}^{\infty}\frac{dk}{2\pi^{2}}\ \frac{k^{2}}{\sqrt{k^{2}+M^{2}}}%
\frac{1}{e^{\sqrt{k^{2}+M^{2}}/T}-1}\ ,
\]
\noindent and a divergent vacuum contribution,
\begin{equation}
Q_{V}(M)=\int\frac{d^{3}\vec{k}}{\left(  2\pi\right)  ^{3}}\frac{1}%
{2\sqrt{\vec{k}^{2}+M^{2}}}=\int\frac{d^{4}k}{\left(  2\pi\right)  ^{4}}%
\frac{1}{k^{2}+M^{2}}\text{ .}%
\end{equation}
\noindent For the explicit calculations we need the following integrals:
\begin{gather}
\int\limits_{k}{G}_{\sigma}=Q_{V}(M_{\sigma})+Q_{T}(M_{\sigma})\ ,\\
\int\limits_{k}{G}_{\pi}=\frac{1}{Z^{2}}\left[  Q_{V}\left(\frac{M_{\pi}}%
{Z}\right)+Q_{T}\left(\frac{M_{\pi}}{Z}\right)\right]  \ ,\\
\int\limits_{k}k^{2}{G}_{\pi}=\frac{M_{\pi}^{2}}{Z^{4}}\left[  Q_{V}%
\left(\frac{M_{\pi}}{Z}\right)+Q_{T}\left(\frac{M_{\pi}}{Z}\right)\right]
\label{A5} \ .
\end{gather}
\noindent\newline\ \noindent For the effective potential we need in addition:
\begin{gather}
\frac{1}{2}\int\limits_{k}\left[  D_{\sigma}^{-1}G_{\sigma}-1\right]
=\frac{1}{2}\int\limits_{k}\left[  (-k^{2}+m_{\sigma}^{2})\frac{1}%
{-k^{2}+M_{\sigma}^{2}}-\frac{-k^{2}+M_{\sigma}^{2}}{-k^{2}+M_{\sigma}^{2}%
}\right] \nonumber\\
=\frac{1}{2}(m_{\sigma}^{2}-M_{\sigma}^{2})\int\limits_{k}\frac{1}%
{-k^{2}+M_{\sigma}^{2}}=\frac{1}{2}(m_{\sigma}^{2}-M_{\sigma}^{2})\left[
Q_{V}(M_{\sigma})+Q_{T}(M_{\sigma})\right]  \ ,\\
\frac{1}{2}\int\limits_{k}\left[  D_{\pi}^{-1}G_{\pi}-1\right]  =\frac{1}%
{2}\int\limits_{k}\left[  (-k^{2}+m_{\pi}^{2})\frac{1}{-Z^{2}k^{2}+M_{\pi}%
^{2}}-\frac{-Z^{2}k^{2}+M_{\pi}^{2}}{-Z^{2}k^{2}+M_{\pi}^{2}}\right]
\nonumber\\
=\frac{1}{2}(m_{\pi}^{2}-M_{\pi}^{2})\frac{1}{Z^{2}}\left[  Q_{V}\left(
\frac{M_{\pi}}{Z}\right)  +Q_{T}\left(  \frac{M_{\pi}}{Z}\right)  \right]
+\frac{1}{2}(Z^{2}-1)\frac{M_{\pi}^{2}}{Z^{4}}\left[  Q_{V}\left(
\frac{M_{\pi}}{Z}\right)  +Q_{T}\left(  \frac{M_{\pi}}{Z}\right)  \right]  \ ,
\end{gather}
\noindent%
\begin{gather}
\frac{1}{2}\int\limits_{k}\ln{G_{\sigma}^{-1}}=\frac{1}{2}R_{V}(M_{\sigma
})+\frac{1}{2}R_{T}(M_{\sigma})\ ,\\
\frac{1}{2}\int\limits_{k}\ln{G_{\pi}^{-1}}=\frac{1}{2}R_{V}\left(
\frac{M_{\pi}}{Z}\right)  +\frac{1}{2}R_{T}\left(  \frac{M_{\pi}}{Z}\right)
\ ,
\end{gather}
\noindent
where 
\begin{gather}
R_{T}\left(\frac{M}{Z}\right)
=\frac{T}{\pi^{2}}\int\limits_{0}^{\infty}dkk^{2}\ln\left(
1-e^{-\sqrt{k^{2}+\frac{M^{2}}{Z^{2}}}/T}\right)  \ ,
\end{gather}
\noindent which in the case $M=0$ and $Z\neq 0$ simplifies to
\begin{gather}
R_{T}(0)=-\frac{T^{4}\pi^{2}}{45}\ .
\end{gather}

We note that in expression (\ref{A5}) we have already dropped a
divergent contribution. In Sec.\ \ref{nuts} we justified this omission
within the scope of our work. However, the starting point is the result
\begin{gather}
\label{stp}
 \int_k k^2 G_{\pi} = -\frac{1}{Z^2} T \sum_{n=-\infty}^{\infty} 
\int \frac{d^3 \vec{k}}{(2 \pi)^3} 1 
 +\frac{M_{\pi}^{2}}{Z^{4}}\left[  Q_{V}
   \left(\frac{M_{\pi}}{Z}\right)
+Q_{T}\left(\frac{M_{\pi}}{Z}\right)\right] \ . 
\end{gather}

Except for the discussion in Secs.\ \ref{elina1} and \ref{elina2}, we neglected
the contributions from renormalization, i.e., $Q_{V}\equiv0$ as well as
$R_{V}\equiv0$. This approximation scheme is called the 
trivial regularization.

\section{Dimensional Regularization}
\label{app2}

This appendix contains calculations relevant for the discussion of the
role of the Jacobian and the choice of the regularization scheme.

First we give derivations for the renormalized thermal integrals using
the dimensional regularization scheme. In dimensional regularization
the thermal tadpole integral is given by \cite{Andersen:2004fp}
\begin{gather}
 \int_k \frac{1}{-k^2+M^2} = \frac{1}{(4\pi)^2} \left(\frac{\mu}{M} \right)^{2\epsilon} 
 \left[ T^2 \frac{4 e^{\gamma \epsilon} \Gamma\left(\frac{1}{2} \right)}{\Gamma\left(\frac{5}{2}-1-\epsilon \right)}
 \beta^2 M^{2\epsilon} \int\limits_{0}^{\infty} dk \frac{k^{2-2\epsilon}}{\sqrt{k^2+M^2} \left(e^{\beta \sqrt{k^2+M^2}}-1 \right)}  - \frac{e^{\gamma \epsilon} \Gamma(1+\epsilon)}{\epsilon (1-\epsilon)} M^2  \right] \ ,
\end{gather}
where $\gamma$ is the Euler-Mascheroni constant and $\mu$ denotes an
arbitrary renormalization scale. Expanding the second term about
$\epsilon = 0$ one can isolate the divergent part $-M^2 / (16\pi^2
\epsilon)$. Dropping this divergence (which could be achieved by
introducing appropriate counter terms in the Lagrangian) we obtain in 
the limit $\epsilon \rightarrow 0$
\begin{gather}
 \int_k \frac{1}{-k^2+M^2} = Q_T (M) + Q_{\mu}^{DR} (M)\;,
\end{gather}
 with
\begin{gather} \label{QmuDR}
 Q_{\mu}^{DR} (M) = - \frac{M^2}{16 \pi^2} \left(1+ \ln{\frac{\mu^2}{M^2}} \right) \ .
\end{gather}
Accordingly, using dimensional regularization, we obtain for Eq.\ (\ref{stp})
\begin{gather}
 \int_k k^2 G_{\pi} = \frac{M_{\pi}^{2}}{Z^{4}}\left[  Q_{\mu}^{DR} \left(\frac{M_{\pi}}{Z}\right)+Q_{T}\left(\frac{M_{\pi}}{Z}\right)\right] \ ,
\end{gather}
since the divergent term (\ref{velt}) vanishes in dimensional regularization due to Veltman's rule \cite{Leibbrandt:1975dj}.

\section{Perturbative cancellation of infinities}
\label{cancel}

In polar coordinates, the Jacobian of the 
integration measure leads to the appearance of 
infinite terms $\sim I$, cf.\ Eqs.\ (\ref{Jacobian}) and (\ref{velt}). 
It was shown in Refs.\
\cite{Chervyakov:2000,Argyres:2009em,Argyres:2009sw} that
these terms cancel in perturbation theory. This
cancellation is affected by the momentum-dependent
vertices in the polar Lagrangian (\ref{lpol}). As we will show
in this appendix, such a cancellation does not happen
in a truncation of the CJT effective potential at a given loop-order, i.e.,
in a certain many-body approximation. The reason is, as we shall see in
the following, that
diagrams with momentum-dependent vertices of higher-loop order 
are required to cancel terms $\sim I$ at a lower-loop order.
Expanding the effective potential in a given order in 
the interaction terms $\sim 1/\varphi$ in the Lagrangian (\ref{lpol}), 
we explicitly demonstrate how this cancellation happens in the two lowest
orders. We conjecture (although we cannot prove it) 
that this also works to arbitrarily high order.

Adding the contribution (\ref{Jacobian}) from the Jacobian to
the Lagrangian (\ref{lpol}), performing the shift 
$\sigma \rightarrow \sigma + \phi$, and expanding the 
trigonometric functions as well as the logarithm from the Jacobian
in a power series in the fields, we obtain up to fourth order in the fields
\begin{eqnarray}
{\cal L}_{pol} & = & \frac{1}{2}\,\partial_{\mu}\sigma\partial^{\mu}\sigma
- \frac{1}{2} \left( 6 \lambda \phi^2 - m^2 + I\, \frac{\delta^2}{\phi^2}
\right) \sigma^2 
 +  \frac{1}{2}\, \partial_\mu \pi \partial^\mu \pi - \frac{1}{2}\,
\frac{H \delta^2}{\phi}\, \pi^2 \nonumber \\
& + & \frac{\delta}{\phi}\, \sigma\, \partial_\mu \pi \partial^\mu \pi
+ \frac{\delta^2}{2 \phi^2}\, \sigma^2 \partial_\mu \pi \partial^\mu \pi
- 2 \lambda \phi\, \sigma^3 - \frac{\lambda}{2}\, \sigma^4
 +  \frac{H\delta^4}{24 \phi^3}\, \pi^4 - \frac{H \delta^2}{2 \phi^2}\,
\sigma\, \pi^2 \nonumber \\
& + & I\, \frac{\delta}{\phi}\, \sigma + I\, \frac{\delta^3}{3
\phi^3}\, \sigma^3 - I\, \frac{\delta^4}{4 \phi^4}\, \sigma^4 
 -  U(\phi) +O(\sigma^5,\sigma \pi^4) \;. \label{lpol2}
\end{eqnarray}
Here, we introduced a power-counting parameter, $\delta$, in all terms 
arising from the expansions of the transcendental functions in
the Lagrangian (\ref{lpol}), in a way that 
each power of $1/\phi$ is accompanied by a factor $\delta$. Our proof
of cancellation of infinities $\sim I$ will here include terms up to
$O(\delta^3)$. The classical (tree-level) potential is given by
\begin{equation}
U(\phi) = -\frac{m^2}{2}\, \phi^2 + \frac{\lambda}{2}\, \phi^4 - H\, \phi\;.
\end{equation}
The CJT effective potential is still given by Eq.\ (\ref{veffcjt}), 
but the inverse tree-level propagators now read
\begin{eqnarray} \label{tree-levelsigmaprop}
D_\sigma^{-1}(k,\phi) & = & -k^2 - m^2 + 6 \lambda \phi^2 +I\, 
\frac{\delta^2}{\phi^2}\;, \\
D_\pi^{-1}(k,\phi) & = & - k^2 + \frac{H \delta^2}{\phi}\;,
\end{eqnarray}
where apart from the additional factor of $\delta^2$ in the mass
term of the pion propagator,
the tree-level propagator for the sigma
receives an additional contribution from the expansion of
the logarithm arising from the Jacobian, cf.\ Eq.\ (\ref{lpol2}).

\begin{figure}[ptbh]
\includegraphics[scale=0.5]{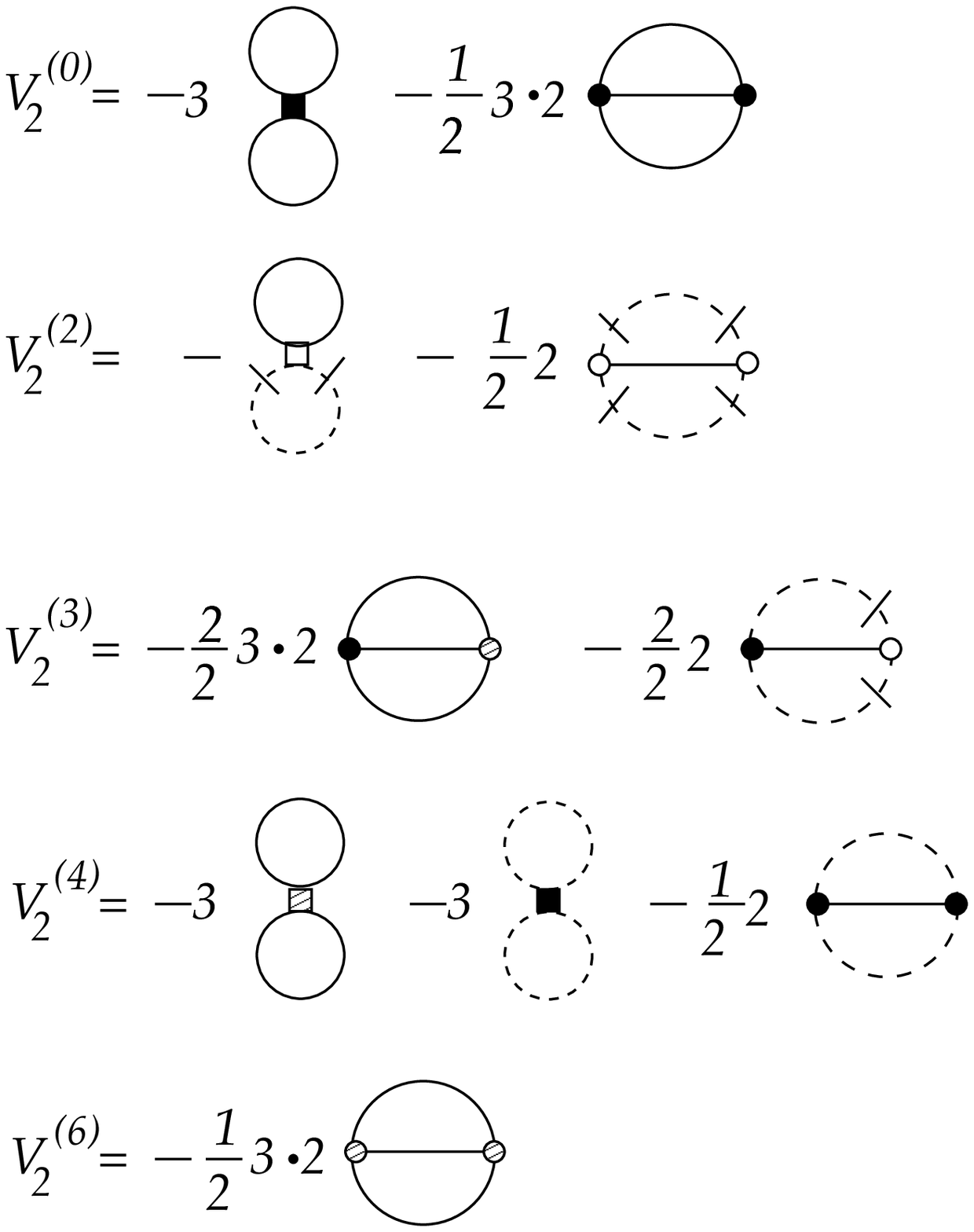}
\caption{2PI Feynman diagrams entering the effective potential
at two-loop order. Solid lines correspond to the sigma, dashed lines to
the pion two-point function. Momentum-dependent
three-point (four-point) vertices are denoted by open circles
(boxes), with additional bars on the attached pion lines.
Regular three- and four-point vertices are denoted by filled
circles and boxes, respectively. Hashed three- and four-point
vertices are proportional to $I$.}
\label{fig2loop}%
\end{figure}

Applying the usual Feynman rules for the construction of the 
2PI contribution to the effective potential (\ref{veffcjt}),
the latter can
now be ordered in powers of $\delta$ accompanying the interaction
terms in Eq.\ (\ref{lpol2}). Up to two-loop order we obtain
\begin{equation}
V_2^{\rm 2-loop}  =  V_2^{(0)} + V_2^{(2)} + V_2^{(3)} + V_2^{(4)}
+V_2^{(6)}\;,
\end{equation}
where
\begin{eqnarray}
V_2^{(0)} & = & \frac{3}{2}\,\lambda \left[ \int_k G_\sigma(k) \right]^2
- 12 \lambda^2 \phi^2 \int_{k,q} G_\sigma(k)\, G_\sigma(q)\, G_\sigma(k+q)\;,
\\
V_2^{(2)} & = & -\, \frac{\delta^2}{2 \phi^2} \int_q G_\sigma(q) \int_k
k^2 G_\pi(k) - \frac{\delta^2}{\phi^2}
\int_{k,q} (k \cdot q)^2\, G_\pi(k)\, G_\pi(q)\, G_\sigma(k+q)\;, 
\label{V_22} \\
V_2^{(3)} & = & 4 \lambda \, I\, \frac{\delta^3}{\phi^2}
\int_{k,q} G_\sigma(k)\, G_\sigma(q)\, G_\sigma(k+q)
- \frac{H \delta^3}{\phi^3} \int_{k,q} k \cdot q\, G_\pi(k)\, G_\pi(q)\,
G_\sigma(k+q)\;, \label{V_23}\\
V_2^{(4)} & = & \frac{3}{4}\, I \, \frac{\delta^4}{\phi^4}
\left[ \int_k G_\sigma(k) \right]^2 - \frac{H \delta^4}{8 \phi^3}
\left[ \int_k G_\pi(k) \right]^2 - \frac{H^2\delta^4}{4 \phi^4}
\int_{k,q} G_\pi(k)\, G_\pi(q)\, G_\sigma(k+q)\;,\\
V_2^{(6)} & = & - I^2\, \frac{\delta^6}{3 \phi^6} 
\int_{k,q} G_\sigma(k)\, G_\sigma(q)\, G_\sigma(k+q)\;.
\end{eqnarray}
Figure \ref{fig2loop} shows these terms in a graphical form, including
combinatorial factors. The minus signs are due to the fact that the effective
potential is proportional to the negative of the pressure.
Terms with two identical (different) vertices are of second order 
in the interaction and thus have an additional factor of 1/2 (2/2).
Other factors are of combinatorial origin and denote the number of
possibilities to connect the vertices with lines.

The stationarity condition of the effective potential with respect to
the one-point function now reads
\begin{eqnarray}
m^2 \varphi - 2 \lambda \varphi^3 + H
& = & \left( 6 \lambda \varphi - I\, \frac{\delta^2}{\varphi^3} \right)
\int_k G_\sigma(k) - \frac{H\delta^2}{2 \varphi^2} \int_k G_\pi(k) \\
& - & 24 \lambda^2 \varphi \int_{k,q} G_\sigma(k)\, G_\sigma(q)\, G_\sigma(k+q)
\nonumber \\
& + & \frac{\delta^2}{\varphi^3} \int_q G_\sigma(q) \int_k k^2 G_\pi(k)
+ 2\, \frac{\delta^2}{\varphi^3} \int_{k,q} (k \cdot q)^2 G_\pi(k)\,
G_\pi(q)\, G_\sigma(k+q) \nonumber \\
& - & 8\, I\, \frac{\lambda \delta^3}{\varphi^3}
\int_{k,q} G_\sigma(k)\, G_\sigma(q)\, G_\sigma(k+q)
+ 3\,\frac{H \delta^3}{\varphi^4}  
\int_{k,q} k \cdot q\, G_\pi(k)\, G_\pi(q)\,G_\sigma(k+q) +O(\delta^4)\;,
\nonumber 
\end{eqnarray}
where the terms in the first line on the right-hand side originate
from the one-loop terms in the effective potential, 
the terms in the second line from $V_2^{(0)}$,  the terms
in the third line from $V_2^{(2)}$, and the terms in the fourth line
from $V_2^{(3)}$. We suppressed terms of higher order,
as our proof of cancellation of infinities extends only up to (and including)
terms of order $O(\delta^3)$. 

The stationarity conditions of the effective potential with respect
to the two-point functions lead to
\begin{eqnarray}
G_\sigma^{-1}(k) & = & - k^2 -m^2 + 6 \lambda \phi^2 
+ I\, \frac{\delta^2}{\phi^2}\nonumber \\
& + & 6 \lambda \int_q G_\sigma(q) - 72\lambda^2 \phi^2 \int_q G_\sigma(q)\,
G_\sigma(k-q) \nonumber \\
& - & \frac{\delta^2}{\phi^2} \int_q q^2 G_\pi(q)
- 2\, \frac{\delta^2}{\phi^2} \int_q [ q \cdot (k-q)]^2G_\pi(q) \,G_\pi(k-q)
\nonumber \\
& + & 24 \,I\, \frac{\lambda \delta^3}{\phi^2} \int_q G_\sigma(q)\,
G_\sigma(k-q) - 2 \,\frac{H \delta^3}{\phi^3}
\int_q  q \cdot (k-q)\,G_\pi(q) \,G_\pi(k-q) + O(\delta^4)\;,
\label{fullsigmaprop} \\
G_\pi^{-1}(k) & = & - k^2  + \frac{H\delta^2}{\phi} \nonumber \\
& - & \frac{\delta^2}{\phi^2}\, k^2  \int_q G_\sigma(q)
- 4\, \frac{\delta^2}{\phi^2} \int_q [k \cdot(q-k)]^2 G_\sigma(q)\,
G_\pi(k-q)\nonumber \\
& - & 4\, \frac{H \delta^3}{\phi^3} \int_q k \cdot (q-k)\, G_\sigma(q)\, 
G_\pi(k-q) + O(\delta^4)\;. \label{fullpionprop}
\end{eqnarray}
At order $O(\delta^2)$, the following infinite terms $\sim I$ arise
in the effective potential:
one in the one-loop term [the second term in Eq.\ (\ref{veffcjt})], 
where the inverse tree-level sigma propagator
appears, which features a term $\sim I \delta^2/\phi^2$, cf.\ Eq.\
(\ref{tree-levelsigmaprop}), and one in each of the two terms from
the two-loop contribution (\ref{V_22}). 
This already demonstrates the above mentioned fact that terms
of higher-loop order are required in order to cancel
infinities at lower-loop order. We now isolate the infinities in the
two-loop terms (\ref{V_22}) and
prove that they cancel against the one from the one-loop term.
To this end, we have to compute these
diagrams explicitly. 

In a perturbative calculation, it is 
sufficient to consider the two-point functions $G_\sigma(k),\,
G_\pi(k)$ appearing in both
the aforementioned one-loop term as well as in the terms in
Eq.\ (\ref{V_22}) only to order $O(\delta^0)$, as each of these terms is
already of order $O(\delta^2)$. To order $O(\delta^0)$,
Eqs.\ (\ref{fullsigmaprop}), (\ref{fullpionprop}) reduce to
\begin{eqnarray} \label{delta2fullpropsigma}
G_\sigma^{-1}(k) & = &  - k^2 -m^2 + 6 \lambda \left[ \phi^2 +  
\int_q G_\sigma(q) \right] - 72\lambda^2 \phi^2 \int_q G_\sigma(q)\,
G_\sigma(k-q) + O(\delta^2)\;,\\
G_\pi^{-1}(k) & = & - k^2 + O(\delta^2)\;. \label{delta2fullproppion}
\end{eqnarray}
Plugging this approximate form of the pion propagator into the 
two terms in the third line of
Eq.\ (\ref{fullsigmaprop}) we obtain after some straightforward steps
\begin{eqnarray}
\lefteqn{ - \frac{\delta^2}{\phi^2} \int_q q^2 G_\pi(q)
- 2\, \frac{\delta^2}{\phi^2} \int_q [ q \cdot (k-q)]^2G_\pi(q) \,G_\pi(k-q) }
\nonumber \\
& = & I\, \frac{\delta^2}{\phi^2} - 2\, \frac{\delta^2}{\phi^2}
\left[ I + \frac{1}{2}\, k^2 \int_q G_\pi(q) + \frac{1}{4}\, (k^2)^2
\int_q G_\pi(q)\, G_\pi(k-q) \right] \nonumber \\
& = & - I\, \frac{\delta^2}{\phi^2}
- k^2 \, \frac{\delta^2}{\phi^2} \int_q G_\pi(q)
- \frac{\delta^2}{2 \phi^2}\, (k^2)^2 \int_q G_\pi(q)\, G_\pi(k-q)\;.
\label{C16}
\end{eqnarray}
The first term on the right-hand side cancels the last term in the
first line of Eq.\ (\ref{fullsigmaprop}), so there are, at least to
order $O(\delta^2)$, no terms
$\sim I$ in the two-point function of the sigma particle. One can
convince oneself that there are also no infinities in the two-point function
of the pion. The same is true for the stationarity condition of the
effective potential with respect to the one-point function.

We now proceed to compute the $O(\delta^2)$ terms in the
effective potential, in the approximation (\ref{delta2fullpropsigma}),
(\ref{delta2fullproppion}) for the sigma and
pion two-point functions, in order to see the cancellation of infinite
terms $\sim I$ between
the one-loop term involving the tree-level sigma propagator and
the two terms of Eq.\ (\ref{V_22}):
\begin{eqnarray}
\frac{1}{2} \int_k D_\sigma^{-1}(k,\phi)\, G_\sigma(k)
+ V_2^{(2)} & = &
\frac{1}{2} \int_k \left(-k^2 - m^2 + 6 \lambda \phi^2 + 
I\,\frac{\delta^2}{\phi^2} \right)\, G_\sigma(k) \nonumber \\
& - & \frac{\delta^2}{2 \phi^2} \int_q G_\sigma(q) \int_k
k^2 G_\pi(k) - \frac{\delta^2}{\phi^2}
\int_{k,q} (k \cdot q)^2\, G_\pi(k)\, G_\pi(q)\, G_\sigma(k+q)\nonumber \\
& = & \frac{1}{2} \int_k \left(-k^2 - m^2 + 6 \lambda \phi^2 \right) 
G_\sigma(k)
+ I\,\frac{\delta^2}{\phi^2} \int_k G_\sigma(k) \nonumber \\
& - & \frac{\delta^2}{ \phi^2}
\int_{k} G_\sigma(k) \int_q [ q \cdot (k-q)]^2 G_\pi(q)\, G_\pi(k-q)\;,
\end{eqnarray}
where we substituted variables $k+q \rightarrow k$ in the last term.
The $q-$integral in this term has already been computed in the above
analysis of the sigma two-point function, cf.\ Eq.\ (\ref{C16}).
Utilizing this, we obtain
\begin{eqnarray}
\frac{1}{2} \int_k D_\sigma^{-1}(k,\phi)\, G_\sigma(k)
+ V_2^{(2)} & = & 
\frac{1}{2} \int_k \left(-k^2 - m^2 + 6 \lambda \phi^2 \right) G_\sigma(k)
+ I\,\frac{\delta^2}{\phi^2} \int_k G_\sigma(k) \nonumber \\
& - & \frac{\delta^2}{ \phi^2}
\int_{k} G_\sigma(k) 
\left[  I + \frac{1}{2}\, k^2 \int_q G_\pi(q) + \frac{1}{4}\, (k^2)^2
\int_q G_\pi(q)\, G_\pi(k-q) \right]\;.
\end{eqnarray}
We observe that the terms $\sim I$ cancel, as claimed.

Now we would like to show the cancellation of infinities up to order 
$O(\delta^3)$ in the effective potential. At two-loop order,
there is one term $\sim I$, cf.\ Eq.\ (\ref{V_23}). As one can 
convince oneself, this infinite contribution cannot be cancelled by
the second term in Eq.\ (\ref{V_23}), which is finite.
The cancellation is, however, affected by a term at the three-loop level.

\begin{figure}[ptbh]
\includegraphics[scale=0.5]{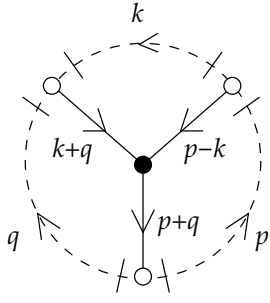}
\caption{The three-loop diagram $D_3$.}
\label{fig3loop}%
\end{figure}

Consider the diagram in Fig.\ \ref{fig3loop}. 
This is of fourth order in interaction vertices, so there is
a factor $1/4!$. There are $3 \cdot 2=6$ possibilities to attach
the sigma lines emerging from the central vertex to the
three other vertices on the pion loop. There are $4 \cdot 2=8$
possibilities to connect the remaining pion lines to form
a diagram with this particular topology. Thus,
\begin{eqnarray}
D_3 & \equiv &  - \frac{1}{4!}\, 6 \cdot 8
\, (-2 \lambda \phi) 
\left( - \frac{\delta}{\phi}\right)^3 
\int_{k,q,p} (k \cdot q) (-k \cdot p) (p \cdot q)\,
G_\pi(k)\, G_\pi(q)\, G_\pi(p)\, G_\sigma(k+q)\, G_\sigma(k-p)\,
G_\sigma(p+q) \nonumber   \\
& = & - \frac{4 \lambda \delta^3}{ \phi^2}
\int_{k,q} G_\sigma(k+q)\, G_\sigma(k)\,G_\sigma(q)
 \int_p[ (k+p) \cdot (q-p)] [ -(k+p) \cdot p]
[p \cdot (q-p)]\,G_\pi(k+p)\, G_\pi(q-p)\, G_\pi(p)\;, \nonumber 
\end{eqnarray}
where we isolated the $p-$integration from the sigma propagators
by the substitutions $k-p \rightarrow k$ and $p+q \rightarrow q$.
The $p-$integral can now be computed, using the approximate form of the
pion two-point function (\ref{delta2fullproppion}),
\begin{eqnarray}
\lefteqn{ \int_p[ (k+p) \cdot (q-p)] [ -(k+p) \cdot p]
[p \cdot (q-p)]\,G_\pi(k+p)\, G_\pi(q-p)\, G_\pi(p)  }
\nonumber \\
& = & \frac{1}{8} \int_p
\left[ (k+q)^2 - (k+p)^2 - (q-p)^2 \right] \left[(k+p)^2 - k^2 + p^2 \right]
\left[q^2 - p^2 - (q-p)^2 \right]\, G_\pi(k+p)\, G_\pi(q-p)\, G_\pi(p) 
\nonumber \\
& = & \frac{1}{8} \int_p \left\{ -k^2 q^2 (k+q)^2 G_\pi(k+p)\, G_\pi(q-p)\,
G_\pi(p) - q^2\left[ (k+q)^2 - (k+p)^2 +k^2  \right]\,G_\pi(q-p)\, G_\pi(p)
\right. \nonumber \\
&  &  - \; k^2 \left[ (k+q)^2 + q^2 - (q-p)^2\right]\,G_\pi(k+p)\, G_\pi(p) 
-(k+q)^2 \left[ k^2 + q^2 - p^2 \right]\, G_\pi(k+p)\, G_\pi(q-p)
\nonumber \\
&  & -\left. 3 \left[ (k+q)^2 +k^2 +q^2 \right] \, G_\pi(p)
+ \left[2 (q-p)^2 + 2(k+p)^2 + (p-k)^2 + (q-p+k)^2 \right]\, 
G_\pi(p) -2  \right\}\;,
\end{eqnarray}
where we made frequent use of substitutions of the integration variable.
Apart from the last term, the only terms which produce infinities are
those where a single pion propagator is accompanied by a factor of $p^2$.
Collecting these, we obtain
\begin{eqnarray}
D_3 &= &- \frac{4 \lambda \delta^3}{ \phi^2} 
\int_{k,q} G_\sigma(k+q)\, G_\sigma(k)\,G_\sigma(q)\,
\left[ I + \mbox{finite}, \, I-\mbox{independent terms} \right]\nonumber\\
& = & - 4 \lambda I \, \frac{\delta^3}{\phi^2}
\int_{k,q} G_\sigma(k+q)\, G_\sigma(k)\,G_\sigma(q)\,
+ \mbox{finite}, \, I-\mbox{independent terms}\;.
\end{eqnarray}
As one can see, this term exactly cancels the infinite term
in Eq.\ (\ref{V_23}).

\bibliographystyle{unsrt}
\bibliography{mybib}


\end{document}